\documentclass[twocolumn,showpacs,preprintnumbers,amsmath,amssymb,pre,longbibliography]{revtex4-1}

\usepackage{color}    
\usepackage{graphicx}
\usepackage{braket} 
\usepackage{dcolumn}
\usepackage{bm}
\usepackage{subfigure}
\usepackage{amssymb}
\usepackage{multirow}
\usepackage{natbib}
    \usepackage{amsmath}
\graphicspath{{plots/}}
\usepackage[english]{babel}

\renewcommand{\vec}[1]{\mathbf{#1}}

\begin{document}


\title{\textit{Ab initio} simulation of warm dense matter}

\author{M.~Bonitz$^1$, T.~Dornheim$^2$, ~Zh.~A.~Moldabekov$^{3,4}$, S.~Zhang$^{1,5}$, P.~Hamann$^1$, H.~K\"ahlert$^1$, A.~Filinov$^1$, K.~Ramakrishna$^{6,7}$, and J.~Vorberger$^6$
}
\affiliation{
 $^1$Institut f\"ur Theoretische Physik und Astrophysik, Christian-Albrechts-Universit\"at zu Kiel,
 Leibnizstra{\ss}e 15, 24098 Kiel, Germany}
 \affiliation{$^2$Center for Advanced Systems Understanding (CASUS), 02826 G\"orlitz, Germany}
 \affiliation{
 $^3$Institute for Experimental and Theoretical Physics, Al-Farabi Kazakh National University, 71 Al-Farabi str.,  
  050040 Almaty, Kazakhstan
  }
\affiliation{$^4$Institute of Applied Sciences and IT, 40-48 Shashkin Str., 050038 Almaty, Kazakhstan}
\affiliation{$^5$Department of Physics, National University of Defense Technology, Changsha, Hunan 410073, China}
\affiliation{$^6$Helmholtz-Zentrum Dresden-Rossendorf, Bautzner Landstra\ss e 400, 01328 Dresden, Germany}
\affiliation{$^7$Technische Universit\"at Dresden, 01062 Dresden, Germany}

\begin{abstract}
Warm dense matter (WDM) -- an exotic state of highly compressed matter --  has attracted high interest in recent years in astrophysics and for dense laboratory systems. At the same time, this state is extremely difficult to treat theoretically. This is due to the simultaneous appearance of quantum degeneracy, Coulomb correlations and thermal effects, as well as the overlap of plasma and condensed phases. Recent breakthroughs are due to the successful application of density functional theory (DFT) methods which, however, often lack the necessary accuracy and predictive capability for WDM applications. The situation has changed with the availability of the first \textit{ab initio} data for the exchange-correlation free energy of the warm dense uniform electron gas (UEG) that were obtained by quantum Monte Carlo (QMC) simulations, for recent reviews, see Dornheim \textit{et al.}, Phys. Plasmas \textbf{24}, 056303 (2017) and Phys. Rep. \textbf{744}, 1-86 (2018). In the present article we review recent further progress in QMC simulations of the warm dense UEG: namely, \textit{ab initio} results for the static local field correction $G(q)$ and for the dynamic structure factor $S(q,\omega)$. These data are of key relevance for the comparison with x-ray scattering experiments at free electron laser facilities and for the improvement of theoretical models.

In the second part of this paper we discuss simulations of WDM out of equilibrium. The theoretical approaches include Born-Oppenheimer molecular dynamics, quantum kinetic theory, time-dependent DFT and hydrodynamics. Here we analyze strengths and limitations of these methods and argue that progress in WDM simulations will require a suitable combination of all methods. A particular role might be played by quantum hydrodynamics, and we concentrate on problems, recent progress, and possible improvements of this method. 
\end{abstract}

\pacs{52.27.Lw, 52.20.-j, 52.40.Hf}
\maketitle

 \section{Introduction}\label{s:intro}
Warm dense matter has become a mature research field on the boarder of plasma physics and condensed matter physics, e.g.~\cite{graziani-book,Fortov2016, moldabekov_pre_18, dornheim_physrep_18}. There are many examples in astrophysics such as the plasma-like matter in brown and white dwarf stars \cite{saumon_the_role_1992, chabrier_quantum_1993,chabrier_cooling_2000}, giant planets, e.g. \cite{schlanges_cpp_95,bezkrovny_pre_4, vorberger_hydrogen-helium_2007, militzer_massive_2008, redmer_icarus_11,nettelmann_saturn_2013}  and the outer crust of neutron stars \cite{Haensel,daligault_electronion_2009}. Warm dense matter is also thought to exist in the interior of our Earth \cite{hausoel_natcom_17}.
In the laboratory, WDM is being routinely produced via laser or ion beam compression or with Z-pinches, see Ref.~\cite{falk_2018} for a recent review article.
Among the facilities we mention the National Ignition facility at Lawrence Livermore National Laboratory \cite{moses_national_2009,hurricane_inertially_2016}, the Z-machine at Sandia National Laboratory \cite{matzen_pulsed-power-driven_2005,knudson_direct_2015}, the Omega laser at the University of Rochester \cite{nora_gigabar_2015},  the Linac Coherent Light Source (LCLS) in Stanford \cite{sperling_free-electron_2015,glenzer_matter_2016}, the European free electron  laser facilities FLASH and X-FEL in Hamburg, Germany \cite{zastrau_resolving_2014,tschentscher_photon_2017},  and the upcoming FAIR facility at GSI Darmstadt, Germany \cite{hoffmann_cpp_18,tahir_cpp19}.  
A particularly exciting application is inertial confinement fusion \cite{moses_national_2009,matzen_pulsed-power-driven_2005,hurricane_inertially_2016} where electronic quantum effects are important during the initial phase. 
Aside from dense plasmas, also many condensed matter systems exhibit WDM behavior -- if they are subject to strong excitation, e.g. by lasers or free electron lasers~\cite{Ernstorfer1033,PhysRevX.6.021003}.

The behavior of all these very diverse systems is characterized, among others, by electronic quantum effects, moderate to strong Coulomb correlations and finite temperature effects. 
Quantum effects of electrons are of relevance at low temperature and/or if matter is very highly compressed, such that the temperature is of the order of (or lower than) the Fermi temperature  (for the relevant parameter range, see Fig.~\ref{fig:0} and, for the parameter definitions, see Sec.~\ref{s:qp_parameters}). 

An important role in the theoretical description of quantum plasmas is being played by quantum kinetic theory
\cite{bohm-pines-3,bohm-pines-4,klimontovich-etal.52a,klimontovich-etal.52b,klimontovich_75,bonitz_aip_12, balzer_pra_10,balzer_pra_10_2}. During the last 25 years, improved and generalized quantum kinetic equations have been derived starting from reduced density operators, e.g. \cite{bogolyubov46,bonitz_qkt}, or nonequilibrium Green functions (NEGF) \cite{kadanoff-baym, keldysh65, balzer-book,schluenzen_19_prl}, for text books see \cite{green-book,bonitz_qkt,blue-book,fortov-book} and references therein. Another direction in quantum plasma theory is first principle computer simulations such as quantum Monte Carlo \cite{sign_cite,militzer_path_2000,filinov_ppcf_01,filinov_pss_00,schoof_cpp11,dornheim_physrep_18,filinov_pre15,schoof_prl15,dornheim_njp15}, semiclassical molecular dynamics with quantum potentials (SC-MD), e.g.~\cite{filinov_pre04}, electronic force fields \cite{goddard_PhysRevLett.99.185003,dai_PhysRevLett.122.015001} and various variants of quantum MD, e.g. \cite{filinov_prb_2, knaup_cpc_02,sjostrom_prl_14,dai_PhysRevLett.104.245001,Kang_2018}.

A recent breakthrough occurred with the application of Kohn-Sham density functional theory (DFT) simulations because they, for the first time, enabled the selfconsistent simulation of realistic warm dense matter, that includes both, plasma and condensed matter phases, e.g. \cite{collins_pre_95,plagemann_njp_2012,witte_prl_17}. Further developments include  orbital-free DFT methods (OF-DFT), e.g. \cite{KARASIEV20143240,dai_doi:10.1016/j.mre.2017.09.001} and  time-dependent DFT (TD-DFT), e.g.  \cite{baczewski_prl_16}. 
In DFT simulations, however, a bottleneck  is the exchange-correlation (XC) functional for which a variety of options exist, the accuracy of which is often poorly known, what limits the predictive power of the method. This requires tests against independent methods such as quantum Monte Carlo simulations for the electron component \cite{dornheim_physrep_18} or  against electron-ion quantum Monte Carlo \cite{pierleoni_cpp19,PhysRevB.89.184106,PhysRevB.93.035121}. Also, the use of finite-temperature functionals was shown to be important \cite{groth_prl17,karasiev_PhysRevB.99.195134} when the XC-contribution is comparable to the thermal energy, see Ref.~\cite{karasiev_2016} for a topical discussion and Ref.~\cite{ramakrishna2020influence} for an extensive investigation of hydrogen.
One goal of this paper is to present an overview of these results and discuss future research directions.

Motivated by time-resolved experiments, e.g. \cite{faeustlin_prl10},  the theoretical description of the
nonequilibrium dynamics of warm dense matter is attracting increasing interest, e.g. \cite{rethfeld_cpp19}.
Time-dependent x-ray Thomson scattering was modelled in Refs.~\cite{gericke_prl11, bornath_cpp19}. Here, a powerful method are quantum kinetic equations \cite{kosse-etal.97,kremp_99_pre} and nonequilibrium Green functions, e.g.~\cite{semkat_99_pre,bonitz_pss_18}.

All of the above mentioned simulation approaches are complex and require substantial amounts of computer time. At the same time, the above mentioned simulations are currently only feasible for small length scales and simulation durations. Therefore, simplified models that would allow to reach larger length and time scales are highly desirable. A possible candidate are fluid models for quantum 
plasmas that are obtained via a suitable coarse graining procedure, as in the case of classical plasmas.
Quantum hydrodynamics (QHD) models for dense plasmas have experienced high activity since the work of Manfredi and Haas  \cite{manfredi_prb_01,manfredi_fields_05}. However, their version of QHD involved several assumptions the validity of which remained open for a long time. Corrections of the coefficients in the QHD equations were recently obtained in Ref.~\cite{michta_cpp15,zhandos_pop18}, and a systematic derivation of the QHD equations from the time-dependent Kohn-Sham equations was given in Ref.~\cite{bonitz_pop_19}. We also mention a recent alternative approach that is based on the computation of semiclassical Bohm trajectories \cite{gregori_bohm_18}.

The goal of this paper is to present a summary of some of the recent in \textit{ab initio} simulations of the electron gas under warm dense matter conditions, including thermodynamic functions and local field corrections developments. Furthermore, we summarize recent progress in the field of QHD for quantum plasmas. In addition to an overview of recent developments, we present new results for a) the application of the finite-temperature exchange correlation free energy in DFT simulations of dense hydrogen and carbon, Sec.~\ref{s:dft}, b) for the dynamic density response function, $\chi(\omega,q)$, Sec.~\ref{ss:dsf}, c) for the screened potential of an ion in a correlated plasma, based on \textit{ab initio} QMC input for the local field correction, Sec.~\ref{ss:qhd-discussion} and d) on the dispersion of ion-acoustic modes in a correlated quantum plasma, Sec.~\ref{ss:ion-acoustic}.

 This paper is organized as follows: in Sec.~\ref{s:qp_parameters} we recall the main parameters of warm dense matter and the relevant temperature and density range. 
Section \ref{s:qmc} presents an overview on recent quantum Monte Carlo simulations followed by finite-temperature DFT results in Sec.~\ref{s:dft}. WDM out of equilibrium and its treatment via a QHD model is discussed in Sec.~\ref{s:nonequilibrium}. 

  \section{Warm dense matter parameters}\label{s:qp_parameters}
\begin{figure}[t]
\includegraphics[width=0.48\textwidth]{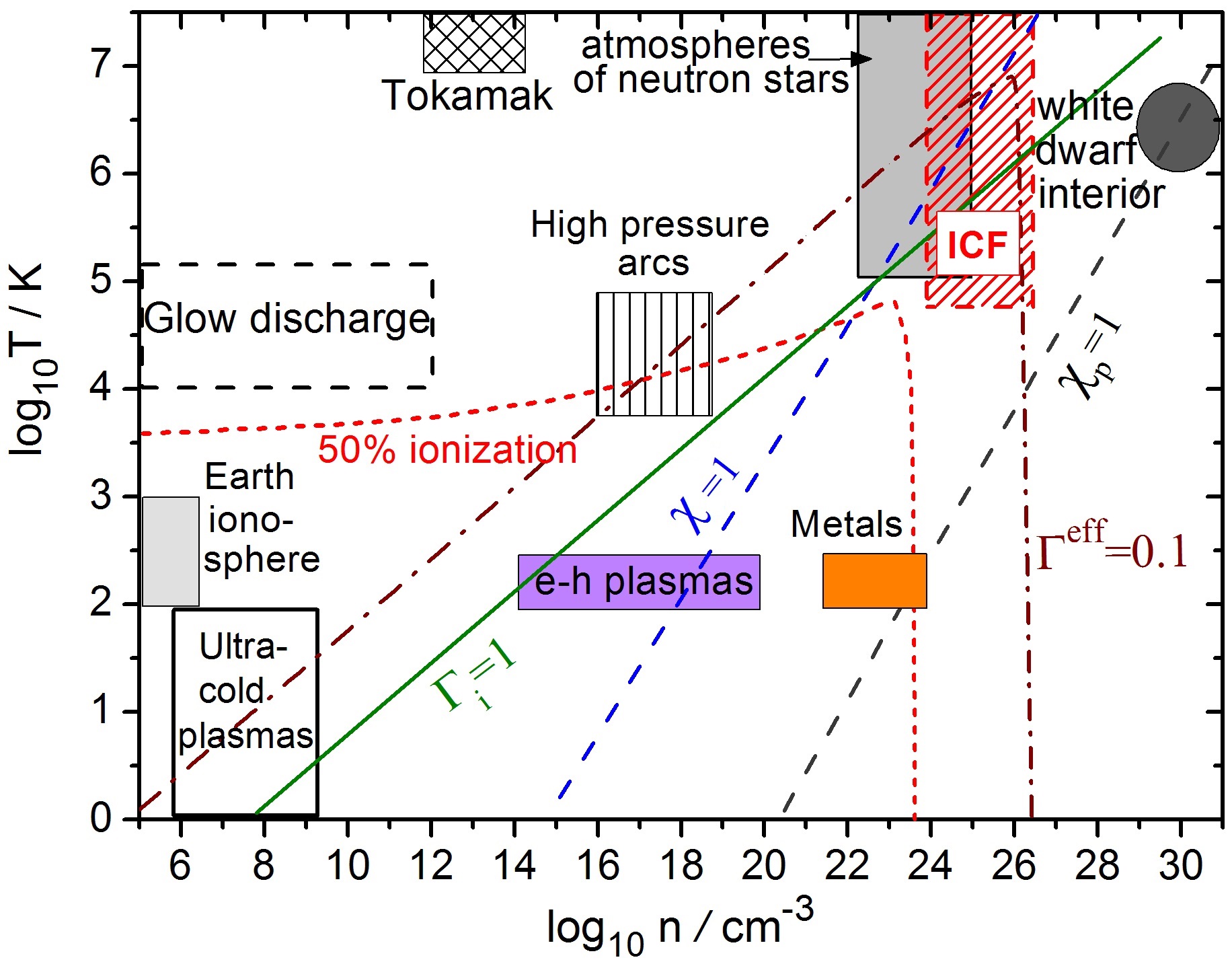}
\caption{Density-temperature plane with examples of plasmas and characteristic plasma parameters. ICF  denotes inertial confinement fusion. Metals (semiconductors) refers to the electron gas in metals (electron-hole plasma in semiconductors). Weak electronic coupling is found outside the line $\Gamma^{\rm eff}=0.1$, cf. Eq.~(\ref{eq:gamma-eff}). Electronic (ionic) \textit{quantum effects} are observed to the right of the line  $\chi = 1$ ($\chi_{\rm p} = 1$). The coupling strength of quantum electrons increases with $r_s$ (with decreasing density). Atomic ionization due to thermal effects (due to pressure ionization) is dominant above (to the right of) the red line, $\alpha^{ion}=0.5$,  for the case of an equilibrium hydrogen plasma~\cite{Sheffield}). The values of $\chi_p$ and $r_s$ refer to the case of hydrogen. Figure modified from Ref.~\cite{bonitz_pop_19}.
}
\label{fig:0}
\end{figure}

Let us recall the basic parameters of warm dense matter \cite{bonitz_qkt,bonitz_pop_19}: 
the first are the electron degeneracy parameters $\theta = k_B T /E_{F} $ and 
$\chi=n\Lambda^3$, where $ \Lambda=h/\sqrt{2\pi m k_B T}$, is 
the thermal DeBroglie wave length, and the Fermi energy of electrons (in 3D) is 
\begin{equation}
    E_{F} = \frac{\hbar^2}{2m}(3\pi^2 n)^{2/3}\,,
    \label{eq:ef-def}
\end{equation}
where $n$ is the electron density and $T$ the electron temperature.
The ion degeneracy parameter 
is a factor $(m T/m_iT_i)^{3/2}$ smaller than the one of the electrons and typically negligible for WDM.
Second is the classical coupling parameter of ions $\Gamma_i= Q_i^2/(a_ik_BT_i)$, where $Q_i$ is the ion charge, and $a_i$ is the mean inter-ionic distance.
Further, the quantum coupling parameter (Brueckner parameter) of electrons in the low-temperature limit is,
    \begin{equation}
    r_s = \frac{a}{a_B}, \qquad a_B = \frac{\hbar^2\epsilon_b}{m_r Q_i e},  
    \label{eq:rs-def}
    \end{equation}
    where $a=(4/3\pi n)^{-1/3}$ denotes the mean  distance between two electrons,  $a_B$ is the Bohr radius, and $m_r=m m_i/(m+m_i)$ and $\epsilon_b$ are the reduced mass and background dielectric constant, respectively, for hydrogen $m_r \approx m\;$, $\epsilon_b=1$, and $a_B=0.529\AA$. Note that another way to measure the coupling strength in the degenerate limit, that is directly related to $r_s$, is via 
    \begin{align}
        \Gamma^2_q = \frac{(\hbar \omega_{pl})^2}{E^2_F} = r_s \cdot \frac{16}{9\pi}\left(\frac{12}{\pi} \right)^{1/3} \approx 0.88 \cdot r_s \,. 
    \end{align}
We can introduce an effective coupling parameter that interpolates between the classical and strongly degenerate limits,
\begin{align}
    \Gamma^{\rm eff} = \frac{e^2/a}{\left[(k_BT)^2+E_F^2\right]^{1/2} } = \frac{e^2}{ak_BT}\frac{1}{\left( 1 + \Theta^{-2} \right)^{1/2}}\,,
    \label{eq:gamma-eff}
\end{align}
and a simple estimate for the boundary between ideal and nonideal plasmas is the line $\Gamma^{\rm eff} =0.1$ that has been included in Fig.~\ref{fig:0}.
Finally, the degree of ionization of the plasma -- the ratio of the number of free electrons to the total (free plus bound) electron number, $\alpha^{ion}=n/n_{tot}$, determines how relevant plasma properties are compared to neutral gas or fluid effects.
%
\begin{figure}[t]
\includegraphics[width=0.52\textwidth]{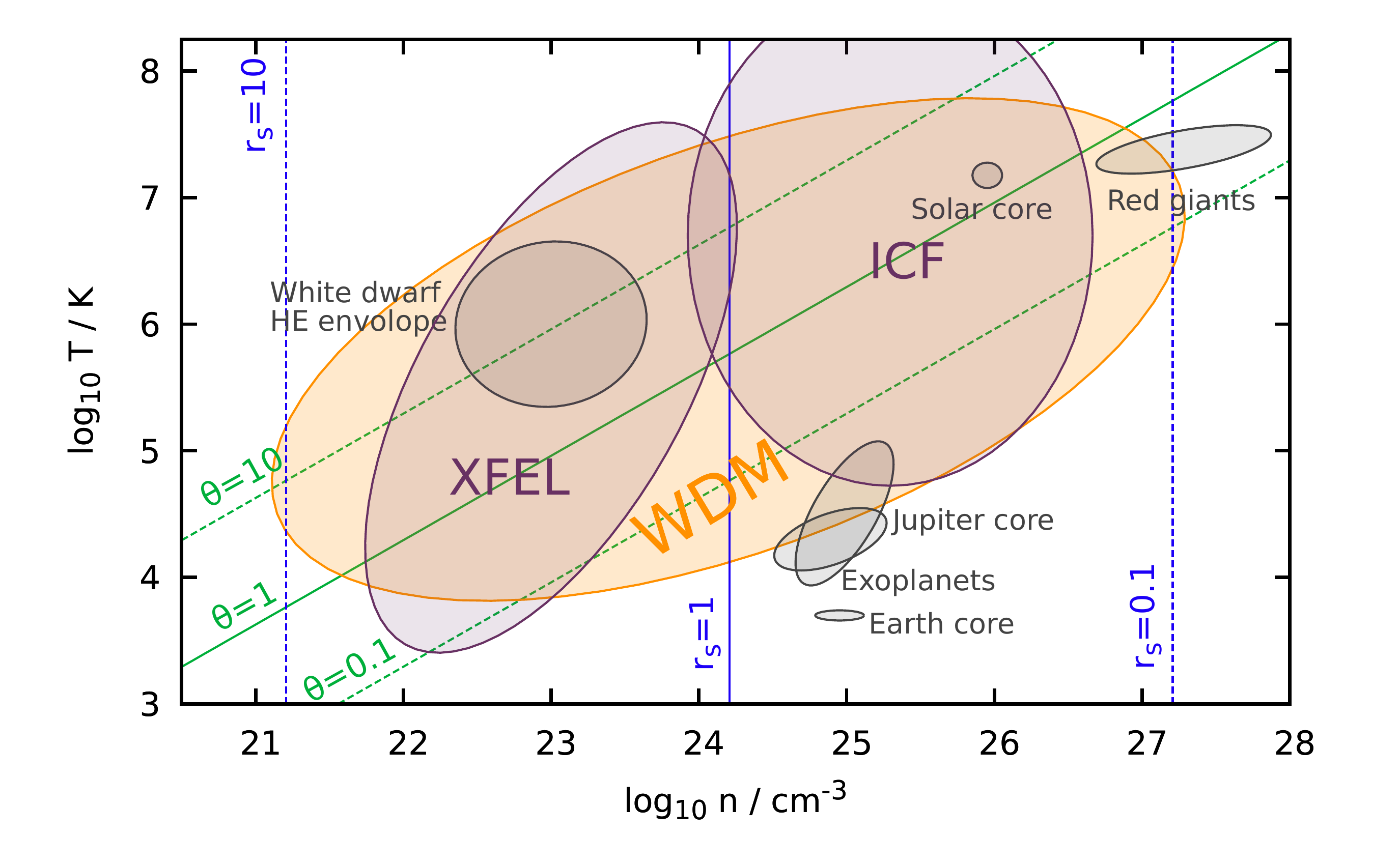}
\caption{Density-temperature plane around WDM parameters with a few relevant examples.
Electronic  \textit{quantum effects} are observed for  $\Theta \lesssim 1$. The coupling strength of quantum electrons increases with $r_s$ (with decreasing density). Note that the values of $\Theta$ and $r_s$ refer to jellium (electrons in fully ionized hydrogen). Adapted from Ref.~\cite{dornheim_physrep_18}.
}
\label{fig:overview}
\end{figure}

The parameters $\chi, \chi_i$ and $\Gamma_i$ are shown in Fig.~\ref{fig:0} where we indicate where these parameters equal one.   
Note that the classical coupling parameter increases with density whereas the quantum coupling parameter decreases with the density $n$. We underline that the parameters $a_B, E_{F}, \Lambda, \theta$ contain the density of free electrons and $\Gamma_i, \chi_i$ the density of free ions. This means the lines of constant $\Gamma_i, \chi, \chi_i, r_s$ shown in Fig.~\ref{fig:0} refer to the free electron (ion) density. In cases when the plasma is only partially ionized the free electron density has to be replaced by $n\to \alpha^{ion}\times n$. 
The degree of ionization decreases when the temperature is lowered, according to the Saha equation, $\alpha^{ion} \sim e^{-|E_b|/k_BT}$, where $E_b$ denotes the binding energy of the atom, and in Fig.~\ref{fig:0} we indicate the line where a classical hydrogen plasma has a degree of ionization of $0.5$. Qualitatively, a quantum plasma is found to the right of this line. Figure~\ref{fig:overview} shows a zoom into the warm dense matter range and also contains lines of constant $r_s$- and $\Theta$-values.

\section{Quantum Monte Carlo Simulations of the Uniform Electron Gas}\label{s:qmc}
\subsection{Summary of \textit{ab initio} static results\label{sec:static}}
The uniform electron gas (UEG) is one of the most fundamental model systems in physics~\cite{loos_gill_2016,giuliani2005quantum,dornheim_physrep_18}. In particular, the accurate parametrization~\cite{vosko_wilk_nusair,perdew_zunger} of the ground-state exchange--correlation energy $e_\textnormal{xc}(r_s)$, based on \textit{ab inito} quantum Monte Carlo simulations~\cite{ceperley_alder}, has been essential for the striking success of density functional theory. While the influence of temperature on the electrons is negligible for most applications in, e.g., condensed matter or chemistry, the recent interest in matter under extreme conditions has led to new demands regarding our understanding of the UEG. More specifically, it has long been known that a thermal DFT~\cite{mermin_1965,gupta_rajagopal_1982} simulation of warm dense matter, see Sec.~\ref{s:dft}, requires a parametrization of the exchange-correlation free energy $f_\textnormal{xc}(r_s,\theta)$, which explicitly depends both on density and on the temperature~\cite{computation4020016,karasiev_2016}.

Consequently, many such parametrizations have been presented over the last decades that are based on various approximations such as dielectric theories~\cite{Perrot_Dharma_1984,Ichimaru_1987,Tanaka_1985,sjostrom_dufty_2013,Tanaka_2016,Tanaka_2017}, quantum--classical mappings~\cite{Perrot_Dharma_2000,Liu_Wu_2014}, and perturbative expansions~\cite{Ebeling_1982,Ebeling_1985}, see Refs.~\cite{groth_cpp17,dornheim_physrep_18} for a topical overview.
In addition, Brown \textit{et al.}~\cite{Brown_2014} presented the first quantum Monte Carlo results for the warm dense UEG using the restricted path integral Monte Carlo (RPIMC) method, which have subsequently been used as input for many applications~\cite{Sjostrom_Gradient_2014,Brown_PRB_2013,Karasiev_PRL_2018}, most notably the parametrization of $f_\textnormal{xc}$ by Karasiev \textit{et al.}~\cite{ksdt} (KSDT).

While being an important mile stone, these data have been obtained on the basis of the uncontrolled fixed node approximation~\cite{Ceperley1991}, which has recently been revealed to be surprisingly inaccurate with systematic errors in the exchange--correlation energy exceeding $10\%$, at high density and low temperature~\cite{schoof_prl15}.
This unsatisfactory situation has sparked a surge of new developments in the field of fermionic QMC simulations at finite temperature~\cite{overcoming,Malone_PRL_2016,Malone_JCP_2015,Blunt_PRB_2014,dornheim_njp15,schoof_cpp15,dornheim_jcp15,dornheim_jcp_19,dornheim_cpp_19}. In particular, Groth, Dornheim, and co-workers have introduced a combination of two complementary QMC methods---permutation blocking PIMC (PB-PIMC) and configuration PIMC (CPIMC)---that allows for a highly accurate description of the UEG over a broad parameter range without the fixed node approximation. After developing a new finite-size correction scheme~\cite{dornheim_prl16}, the same authors presented the first \textit{ab initio} parametrization of $f_\textnormal{xc}$ with respect to density, temperature, and spin-polarization covering the entire WDM regime with an unprecedented accuracy of $\sim0.3\%$.

\begin{figure*}
    \centering
\includegraphics[width=0.75\textwidth]{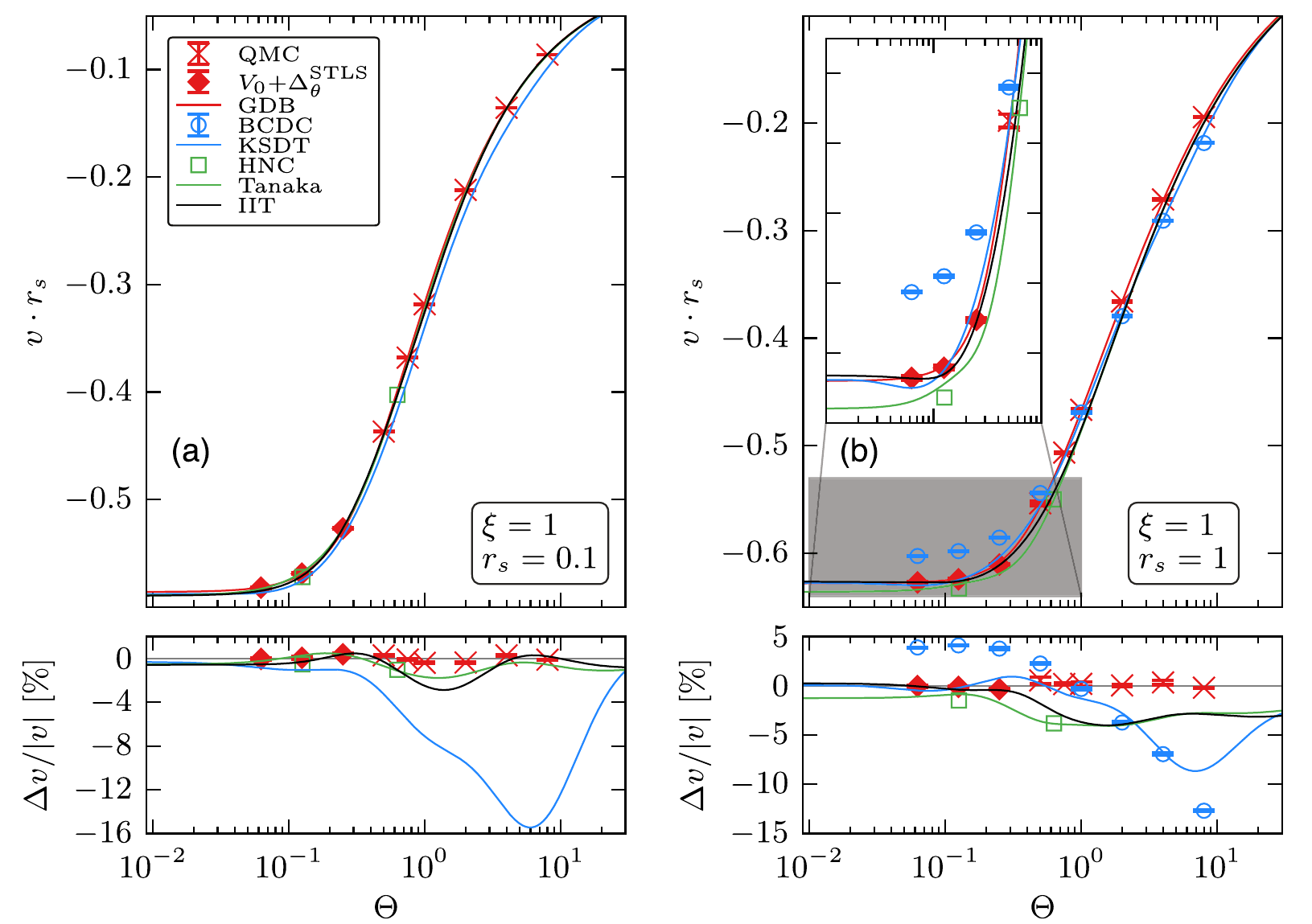}
    \caption{Temperature dependence of the interaction energy of the spin-polarized ($\xi=1$) UEG [the polarization is defined as $\xi=(N^\uparrow-N^\downarrow)/N$] for two densities, computed from different methods and parametrizations. Red crosses: \textit{ab initio} QMC data, red line: parametrization of the QMC data by Groth \textit{et al.} (GDB)~\cite{groth_prl17}. Blue circles: extrapolated RPIMC data by Brown \textit{et al.} (BCDC)~\cite{Brown_2014}. Blue line: RPIMC-based parametrization by Karasiev \textit{et al.} (KSDT)~\cite{ksdt}. Green squares and line: HNC-based dielectric data points and parametrization thereof by Tanaka~\cite{Tanaka_2016}. Black line: STLS-based parametrization by Ichimaru \textit{et al.} (IIT)~\cite{Ichimaru_1987}. The bottom panel shows the relative deviation in $v$ with respect to the GDB reference data.  Taken from Ref.~\cite{dornheim_physrep_18} with the permission of the authors.}
    \label{fig:fxc_comparison}
\end{figure*}

This is illustrated in Fig.~\ref{fig:fxc_comparison}, where we show the temperature-dependence of the interaction energy of the spin-polarized UEG for two different densities and compare different parametrizations and data sets. The red solid line corresponds to the parametrization by Groth \textit{et al.}~\cite{groth_prl17} (GDB, referred to as \textit{GDSMFB} hereafter), which is based on finite-$T$ QMC data (red crosses) for $0.5 \leq \theta \leq 8 $ and temperature-corrected ground-state QMC data (red diamonds) for $\theta\leq0.25$. The other curves depict the RPIMC data from Ref.~\cite{Brown_2014} (BCDC, blue circles) and a corresponding parametrization (KSDT, solid blue), HNC data and a parametrization thereof by Tanaka~\cite{Tanaka_2016} (HNC, green squares and line), and the STLS-based parametrization by Ichimaru \textit{et al}.~\cite{Ichimaru_1987}. At high density ($r_s=0.1$, panel a), both dielectric theories agree relatively well with the GDSMFB reference data (see also the bottom panel showing the relative deviation), while the KSDT curve exhibits deviations of up to $\Delta v/v\sim15\%$. This is a direct consequence of the absence of RPIMC input data in this regime, and the insufficient finite-size correction of these data, where they are available~\cite{dornheim_prl16,dornheim_pop17,dornheim_physrep_18}. In the WDM regime ($r_s=1$, panel b), the situation somewhat changes as both, HNC and STLS, become less accurate and exhibit deviations of up to $\Delta v/v\sim 5\%$ in the relevant temperature range. Moreover, the RPIMC data exhibit systematic deviations from the other curves as they are systematically too large, for small $\theta$, and too low, in the opposite case. This is due to a combination of the fixed-node approximation and the extrapolation to the thermodynamic limit, see Ref.~\cite{dornheim_physrep_18} for an extensive discussion. Interestingly, the KSDT curve is remarkably accurate in the low-temperature limit and does not reproduce the biased RPIMC input data on which it is based. Still, there occur deviations of up to $\Delta v/v\sim8\%$, at elevated temperature.

In the mean time, the availability of the accurate GDSMFB benchmark data has led to a revised version of the KSDT parametrization (denoted as \textit{corrKSDT} in Refs.~\cite{karasiev_gga_18,karasiev_PhysRevB.99.195134}), which basically reproduces GDSMFB over the entire WDM regime~\cite{karasiev_PhysRevB.99.195134}. First and foremost, we note that both GDSMFB and corrKSDT are suitable to be used as an exchange--correlation functional on the level of the local density approximation~\cite{ramakrishna2020influence} [the GDSMFB parametrization is available in the libxc library as ``GDSMFB'', cf. Sec.~\ref{s:dft}], and as the basis for more sophisticated functionals such as a temperature-dependent generalized gradient approximation~\cite{karasiev_gga_18}. This opens up new avenues for DFT simulations of WDM systems without neglecting thermal effects in the XC-functional itself.
On the other hand, Karasiev \textit{et al.}~\cite{karasiev_PhysRevB.99.195134} have found that there occur some potentially unphysical oscillations in quantities, that are derived from $f_\textnormal{xc}$, such as the specific heat $C_V$. This is not surprising as $C_V$ involves the second derivative of the fit with respect to the temperature which may contain a large error. In addition, the entropy was found to become negative at strong coupling and low temperature, which, however, is outside of the intended domains of application of both GDSMFB and corrKSDT.

Let us conclude this section by proposing a few possible solutions to the remaining open questions: i) the unphysical behaviour in the second- and higher-order derivatives are most likely a consequence of the functional form of the GDSMFB and corrKSDT parametrizations~\cite{karasiev_PhysRevB.99.195134}. Therefore, addressing this issue would require a modification of the corresponding Pad\'{e} approximation to automatically fulfill some additional constraints. ii) the current validity domain of, e.g., the GDSMFBB parametrization to $0\leq r_s \leq 20$ can be significantly extended by incorporating the recent \textit{ab initio} PIMC results for the electron liquid regime ($20\leq r_s\leq100$) by Dornheim \textit{et al.}~\cite{dornheim2019strongly}. iii) new \textit{ab intio} QMC results at low temperature, $\theta \sim 10^{-1}$, could help to more accurately resolve open thermodynamic questions like the effective mass enhancement~\cite{Eich_PRB_2017}, but are difficult to obtain due to the notorious fermion sign problem~\cite{dornheim_pre_2019}.  iv) Neural networks are known to be valuable as universal function approximators (see also Sec.~\ref{sec:LFC}), and can be designed to fulfill all known constraints on the UEG. Moreover, they constitute a handy way to combine data for different quantities from different methods into a single, unified representation.

\subsection{Summary of \textit{ab initio} results for the static local field correction\label{sec:LFC}}
One important step in going beyond local approximations, such as LDA or GGAs, is to consider the response of an electron gas to an external harmonic perturbation [cf.~Eq.~(\ref{eq:A})], which is described by the density response function
\begin{eqnarray}\label{eq:define_LFC}
\chi({q},\omega) = \frac{ \chi_0({q},\omega) }{ 1 - \tilde v_q\big[1-G({q},\omega)\big]\chi_0({q},\omega)} \quad ,
\end{eqnarray}
with $q$ and $\omega$ being the corresponding wave number and frequency, respectively, and $\tilde v_q=4\pi e^2/q^2$ [frequently atomic units are used, then this becomes $4\pi/q^2$] is the Fourier transform of the Coulomb potential. Further, $\chi_0(q,\omega)$ denotes the usual Lindhard function that describes the density response of the ideal (i.e., noninteracting system)~\cite{giuliani2005quantum}, and the local field correction $G(q,\omega)$ entails the full frequency- and wave number resolved information about exchange--correlation effects on $\chi$~\cite{Kugler1975}. For example, setting $G(q,\omega)=0$ in Eq.~(\ref{eq:define_LFC}) leads to the widely used random phase approximation (RPA), which describes the density response of the electron gas on a mean field level. 

Consequently, the LFC is of paramount importance to incorporate nonlocal exchange--correlation effects into other theories, like QHD~\cite{michta_cpp15,zhandos_pop18,Diaw2017}, effective potentials~\cite{Senatore_effective_potential_1996,moldabekov_pre15,moldabekov_pop15,zhandos_cpp17,moldabekov_pre_18, zhandos_cpp16, zhandos_wake_cpp19}, and the construction of advanced exchange--correlation functionals for DFT~\cite{Lu_2014,Patrick_2015,Pribram_Jones_PRL_2016,Olsen2019} and time-dependent DFT~\cite{baczewski_prl_16}. Moreover, it can directly be used to compute important material properties like the stopping power~\cite{Nagy1985,Stopping_Power_2017,Cayzac2017}, electrical and thermal conductivities~\cite{Veysman_PRE_2016,Desjarlais_PRE_2017}, and energy transfer rates~\cite{Vorberger_PRE_2010}. Finally, we mention the interpretation of XRTS experiments~\cite{redmer_glenzer_2009,Plagemann_2012,PhysRevE.81.026405,PhysRevLett.105.075003,Kraus_2018}, e.g., within the Chihara decomposition~\cite{Chihara_1987}, which is of paramount importance as a method of diagnostics.

Naturally, there have been many attempts to find suitable approximations for $G(q,\omega)$, most commonly within the purview of dielectric theories~\cite{PhysRevB.35.2720,PhysRevB.6.875,doi:10.1143/JPSJ.55.2278,PhysRev.176.589,PhysRevA.26.603}. The first accurate benchmark data for the LFC have been obtained by Moroni \textit{et al.}~\cite{Moroni_PRL_1992,Moroni_PRL_1995} by performing ground-state QMC simulations of a perturbed electron gas governed by the Hamiltonian
\begin{eqnarray}\label{eq:A}
\hat H = \hat H_\textnormal{UEG} + 2A \sum_{k=1}^N \textnormal{cos}( \hat{\mathbf{r}}_k \cdot \hat{\mathbf{q}} ) \quad ,
\end{eqnarray}
with $\hat H_\textnormal{UEG}$ corresponding to the unperturbed UEG, and $A$ being the perturbation amplitude. More specifically, they have performed multiple simulations for a single wave number $q$ to measure the response of the electron gas in dependence of $A$, which is linear for small $A$ with $\chi(q,0)$ being the slope. While being limited to the static limit (i.e., $\omega=0$), these data have subsequently been used as input for the parametrization of $G(q,0)$
by Corradini \textit{et al.}~\cite{Corradini_PRB_1998} (CDOP), which, in turn, has been used for many applications, e.g., Refs.~\cite{RevModPhys.74.601,PhysRevB.70.205107,Lu_2014,Patrick_2015,zhandos_cpp17,moldabekov_pre_18}.

Recently, Dornheim, Groth and co-workers~\cite{dornheim_pre17,groth_jcp17} have extended the idea behind Eq.~(\ref{eq:A}) to finite temperature using the novel permutation blocking PIMC and configuration PIMC methods, which has allowed to obtain the first \textit{ab initio} results for the static density response of the UEG at WDM conditions. While being conceptually valid and interesting in their own right, these simulations suffer from a prohibitive computational cost: the full characterisation of $G(q,0;r_s,\theta)$ requires a dense grid of densities, temperatures, and wave numbers, $(r_s,\theta,q)$. Unfortunately, each such tuple would, in turn, require multiple simulations for different perturbation amplitudes $A$, and potentially also different $N$ to eliminate possible finite-size effects. Therefore, the aforementioned strategy is valuable to produce accurate benchmark data at specific points, but cannot be feasibly used to generate the bulk of input data needed for a full description of $G(q,0;r_s,\theta)$ covering the entire WDM regime.

A different, more convenient route is given by the imaginary-time version of the fluctuation--dissipation theorem~\cite{Bowen_PRB_1992,dornheim_pre17},
\begin{eqnarray}\label{eq:static_chi}
\chi({q},0) = -n\int_0^\beta \textnormal{d}\tau\ F({q},\tau) \quad ,
\end{eqnarray}
with 
\begin{eqnarray}\label{eq:define_F}
F(q,\tau) = \frac{1}{N}  \braket{\rho(q,\tau)\rho(-q,0)}\,,
\end{eqnarray}
being the density--density correlation function (also known as the intermediate scattering function) evaluated in the imaginary time $\tau\in[0,\beta]$. Equation~(\ref{eq:define_F}) can be straightforwardly computed using the standard PIMC method~\cite{Berne_JCP_1983,Berne1986}, which means that the entire wave-number dependence of $\chi(q,0)$ can be obtained from a single simulation of the unperturbed UEG (i.e., setting $A=0$ in Eq.~[\ref{eq:A})]. The corresponding results for $G(q,0)$ are then computed by solving Eq.~(\ref{eq:define_LFC}) for $G$, i.e.~\cite{groth_prb_19},
\begin{eqnarray}\label{eq:Get_G}
G(q) = 1 - \frac{1}{\tilde v_q}\left( \frac{1}{\chi_0(q)} - \frac{1}{\chi(q)} \right).
\end{eqnarray}
This strategy---in combination with the efficient finite-size correction introduced in Ref.~\cite{groth_jcp17}---was recently used by Dornheim \textit{et al.}~\cite{dornheim_jcp_19-nn} to obtain extensive new \textit{ab initio} PIMC results for $G(q,0)$ for $N_\textnormal{param}\sim50$ density--temperature combinations covering a significant part of the relevant WDM regime. 
These data, together with the CDOP parametrization for $\theta=0$,  were subsequently used as input to train a deep neural net, which takes as input a tuple of $(q,r_s,\theta)$ and predicts as output the corresponding LFC $G(q,0;r_s,\theta)$ in the range of $0\leq \theta \leq 4$, $0.7 \leq r_s \leq 20$, and $0 \leq q \leq 5q_\textnormal{F}$.
\begin{figure}
    \centering
\includegraphics[width=0.465\textwidth]{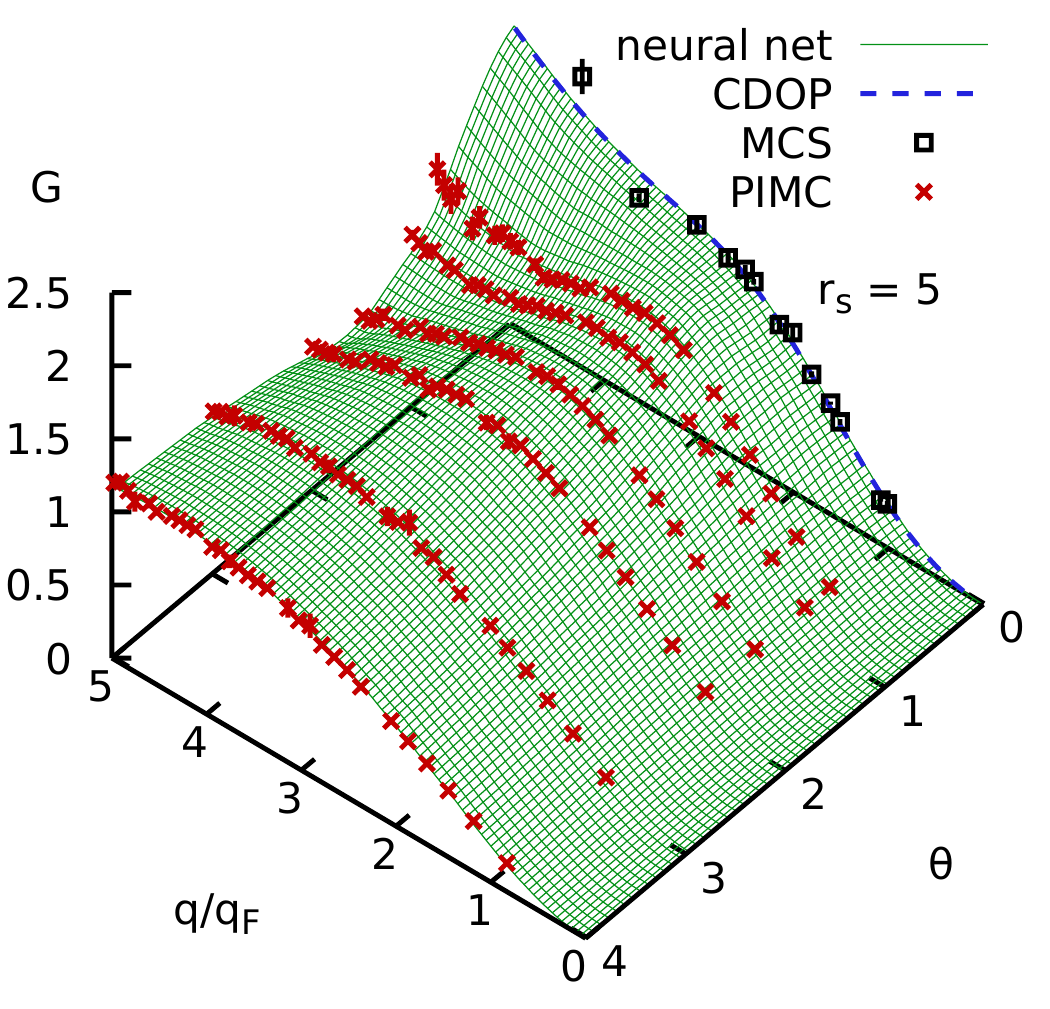}
    \caption{The static local field correction $G(q,0;r_s,\theta)$ in the wave number--temperature plain at $r_s=5$. The black squares and dashed blue line depicts the ground-state QMC data from Ref.~\cite{Moroni_PRL_1995} (MCS) and an accurate parametrization thereof~\cite{Corradini_PRB_1998} (CDOP), and the red crosses correspond to the new finite-temperature PIMC data from Ref.~\cite{dornheim_jcp_19-nn}. The green surface shows the prediction by the neural net, that is available at continuous $q$, $r_s$, and $\theta$.
    Taken from Ref.~\cite{dornheim_jcp_19-nn} with the permission of the authors.
    }
    \label{fig:NN}
\end{figure}

A typical result is shown in Fig.~\ref{fig:NN}, where the static LFC is plotted in the wave number-temperature plain for a fixed value of the density parameter, $r_s=5$. The black squares depict the ground-state QMC results by Moroni \textit{et al.}~\cite{Moroni_PRL_1995}, and the dashed blue line the corresponding CDOP parametrization, which incorporates both, the compressibility sum-rule~\cite{sjostrom_dufty_2013}, for $q\to0$, and the exact large-$q$ limit found by Holas~\cite{Holas1987,Farid_PRB_1993}. The red crosses show the new PIMC data computed from Eq.~(\ref{eq:static_chi}), which is available at a dense grid of wave numbers that is determined by the usual momentum quantization in a finite simulation cell. Finally, the green surface has been evaluated using the neural net published in Ref.~\cite{dornheim_jcp_19-nn}. Evidently, the machine-learning representation nicely reproduces the input data where they are available, and smoothly interpolates in between. Moreover, its capability to predict $G(q,0;r_s,\theta)$ has been validated against independent benchmark data that had not been included in the training procedure, see Ref.~\cite{dornheim_jcp_19-nn} for details.

\begin{figure}
    \centering
\includegraphics[width=0.5\textwidth]{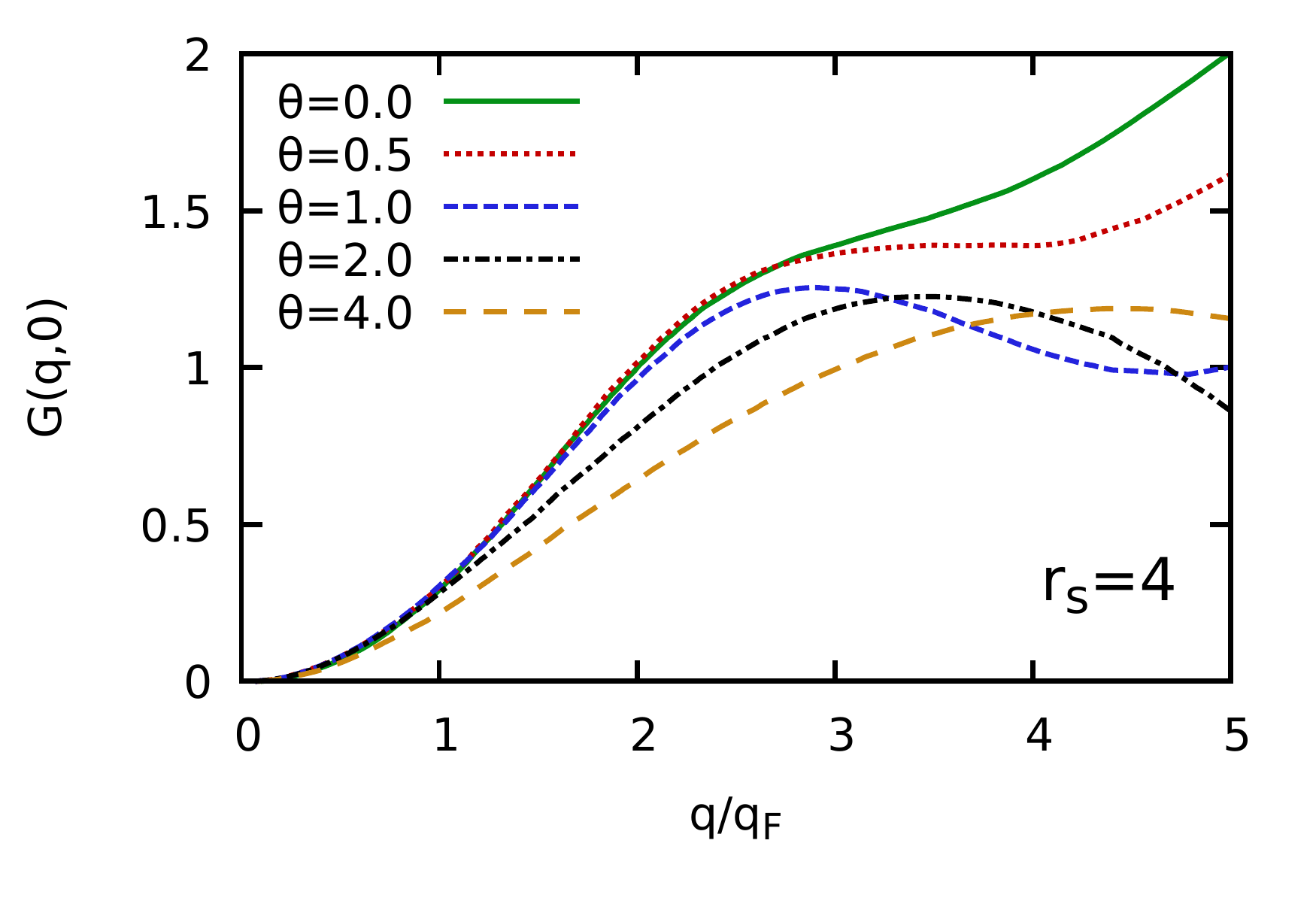}
    \caption{Wave-number dependence of the static LFC at metallic density, $r_s=4$. All curves have been obtained by using the machine-learning representation of $G(q,0;r_s,\theta)$ from Ref.~\cite{dornheim_jcp_19-nn}, which approaches CDOP~\cite{Corradini_PRB_1998} in the low-temperature limit.
    }
    \label{fig:LFC_T}
\end{figure}

Let us conclude this section by explicitly investigating the impact of thermal excitations on the static LFC. To this end, we plot $G$ at a metallic density, $r_s=4$, for five different temperatures in Fig.~\ref{fig:LFC_T}. The solid green curve corresponds to the zero-temperature limit, where $G$ is accurately represented by CDOP. Upon increasing the temperature to $\theta=0.5$ (red dots), the LFC essentially remains unchanged, for $q\lesssim3q_\textnormal{F}$, but exhibits a significant drop and an apparent saddle point, for large wave numbers. At the Fermi temperature (blue dashes), $G(q,0;r_s,\theta)$ exhibits an even more interesting behavior: while it is approximately equal to the $\theta=0$ curve, for $q\lesssim2q_\textnormal{F}$, there appears a complicated shape with a maximum around $q\approx2.7q_\textnormal{F}$, a subsequent minimum at $q\approx4.5q_\textnormal{F}$, and a positive large wave number tail. Finally, further increasing the temperature to $\theta=2$ (dash-dotted black curve) and $\theta=4$ (long-dashed brown curve) leads to significant thermal effects, even for small $q$ values, and $G$ exhibits a pronounced maximum at intermediate wave numbers, followed by a tail with a negative slope~\cite{dornheim_jcp_19-nn}.

For completeness, we mention that new \textit{ab initio} results for $G(q,0;r_s,\theta)$ at strong coupling beyond the WDM regime ($r_s \geq 20$) have recently been presented in Ref.~\cite{dornheim2019strongly}.

\subsection{\textit{Ab initio} dynamic results}\label{ss:dsf}

In the Secs.~\ref{sec:static} and \ref{sec:LFC}, we have outlined the current state of the art regarding both, thermodynamics and the static density response of the UEG in the WDM regime. However, a direct comparison to experiments often requires the calculation of dynamic properties. For example, the central quantity in modern X-ray Thomson scattering (XRTS) experiments~\cite{redmer_glenzer_2009} is given by the dynamic structure factor $S(q,\omega)$, which is defined as the Fourier transform of the intermediate scattering function $F(q,t)$ [cf.~Eq.~(\ref{eq:define_F})],
\begin{eqnarray}
S(q,\omega) = \frac{1}{2\pi} \int_{-\infty}^\infty \textnormal{d}t\ F(q,t) e^{i\omega t} \quad .
\end{eqnarray}
Naturally, the straightforward evaluation of $F(q,t)$ requires real time-dependent simulations~\cite{kwong_prl_00,stefanucci2013nonequilibrium,Kas_PRL_2017}, for which, presently an exact simultaneous treatment of exchange--correlation, thermal, and degeneracy effects is not possible. Therefore, previous results~\cite{nozieres2018theory,Gross_PRL_1985,Dabrowski_PRB_1986,PhysRevB.35.2720,kwong_prl_00,Takada_PRL_2012,Takada_PRB_2016} were based on partly uncontrolled approximations, the quality of which had remained unclear. Moreover, \textit{ab initio} QMC methods, which were pivotal for the accurate description of static properties, as was discussed in the Sections before, are effectively rendered unfeasible regarding time-dependent simulations due to an additional \textit{dynamical sign problem}~\cite{Mak_JCP_1999,PhysRevLett.115.266802}.

An alternative to simulations in real time is given by the method of \textit{analytic continuation}, with the imaginary-time density--density correlation function $F(q,\tau)$, as defined in Eq.~(\ref{eq:define_F}), being the starting point. Recall that the latter can be computed without any approximation from standard PIMC simulations, and it is connected to the dynamic structure factor via a Laplace transform
\begin{eqnarray}\label{eq:FS}
F(\mathbf{q},\tau) = \int_{-\infty}^\infty \textnormal{d}\omega\ S(\mathbf{q},\omega) e^{-\tau\omega} \quad .
\end{eqnarray}
The task at hand is then to find a trial solution $S_\textnormal{trial}(q,\omega)$ that, when being inserted into Eq.~(\ref{eq:FS}), reproduces the PIMC data for $F(q,\tau)$, for all values of $\tau\in[0,\beta]$. Such an inverse Laplace transform is a well-known, but notoriously difficult task, as different $S_\textnormal{trial}(q,\omega)$ might fulfill Eq.~(\ref{eq:FS}) within the given statistical uncertainty~\cite{Jarrell_Review_1996,PhysRevB.94.245140}.
A first step to further restrict the space of possible $S_\textnormal{trial}(q,\omega)$ are frequency moments, which are defined as
\begin{eqnarray}\label{eq:mome}
\braket{\omega^k} = \int_{-\infty}^\infty \textnormal{d}\omega\ S(\mathbf{q},\omega)\ \omega^k \quad ,
\end{eqnarray}
and the results for the cases $k=-1,0,1,3$ are known from different sum-rules~\cite{dornheim_prl_18,groth_prb_19}.
While the combination of Eqs.~(\ref{eq:FS}) and (\ref{eq:mome}) has often turned out to be sufficient to accurately reconstruct $S(q,\omega)$ in the case of, e.g., ultracold bosonic atoms~\cite{Vitali_PRB_2010,filinov_pra_12,Filinov_PRA_2016,Gift2}, we have found that this does not hold for the UEG in the WDM regime, and additional information is indispensible.

To his end, we invoke the fluctuation--dissipation theorem~\cite{giuliani2005quantum}
\begin{eqnarray}\label{eq:FDT}
S(\mathbf{q},\omega) = - \frac{ \textnormal{Im}\chi(\mathbf{q},\omega)  }{ \pi n (1-e^{-\beta\omega})} \quad ,
\end{eqnarray}
which gives a straightforward relation between the DSF and the imaginary part of the dynamic density response function introduced in Sec.~\ref{sec:LFC}. In this way, the original reconstruction problem has been recast into the quest for a suitable dynamic local field correction $G_\textnormal{trial}(q,\omega)$. This is extremely advantageous, as many additional exact constraints on $G$, such as the static and high-frequency limits, and the Kramers-Kronig relation between its real and imaginary part, are known.

In practice, we solve the inversion problem posed by Eq.~(\ref{eq:FS}) by stochastically sampling trial solutions $G_\textnormal{trial}(q,\omega)$ such that a significant number of exact properties are build in by design. Subsequently, we use the corresponding $\chi_\textnormal{trial}(q,\omega)$ to generate trial solutions for the DSF via Eq.~(\ref{eq:FDT}), which are finally plugged into Eqs.~(\ref{eq:FS}) and (\ref{eq:mome}), and discarded if the deviation from our PIMC data is more than the Monte Carlo error bar. The final solution for $S(q,\omega)$ is then computed as the average over a large number of such valid trial solutions, which also allows us to estimate the remaining variance around our estimate for $S(q,\omega)$.

\begin{figure}
    \centering
\includegraphics[width=0.5\textwidth]{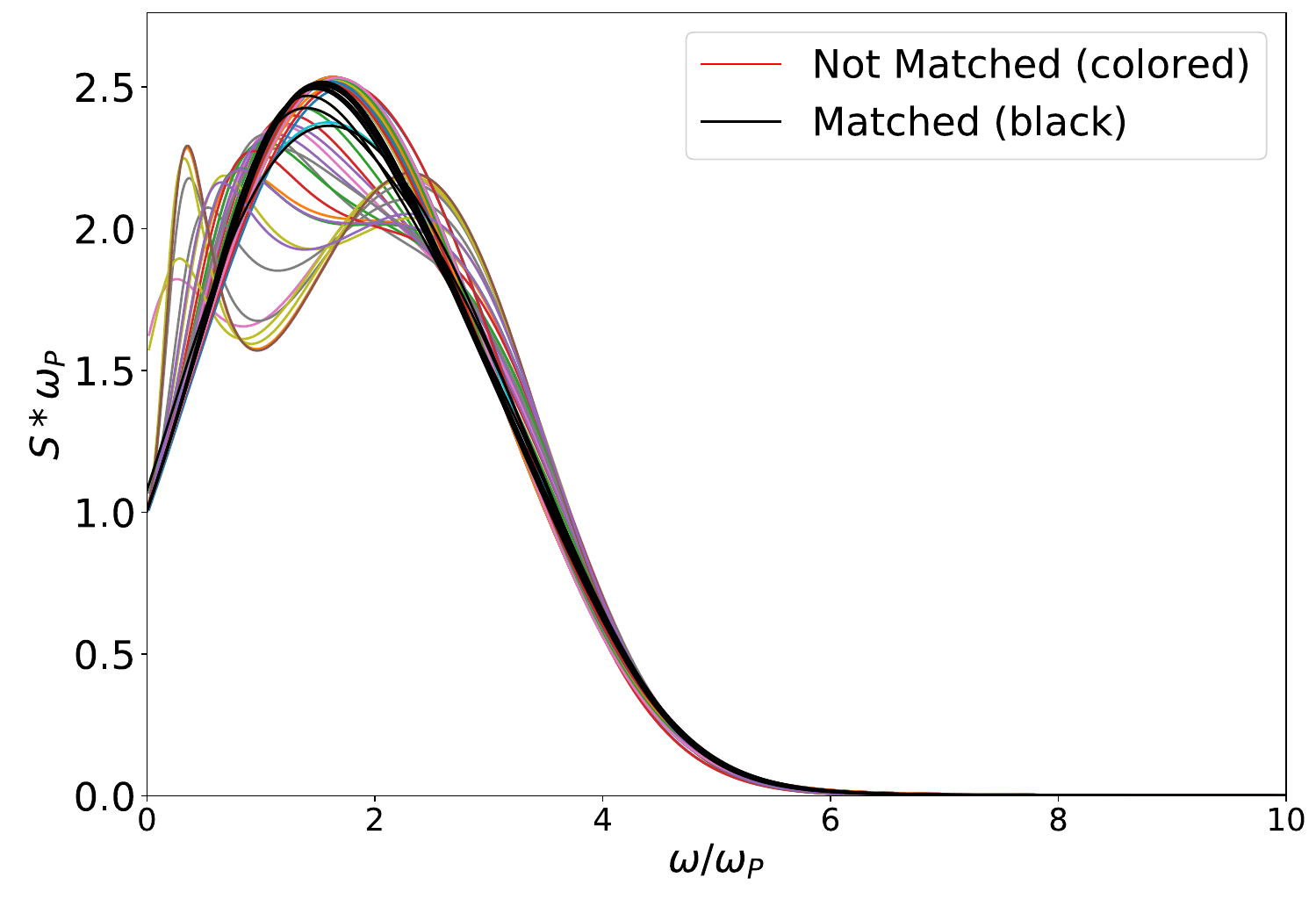}
    \caption{Stochastic sampling method for the dynamic structure factor.
    Shown is the frequency dependence of $S(q,\omega)$ for the unpolarized UEG at $r_s=6$ and $\theta=1$ for a fixed wave number $q\approx2q_\textnormal{F}$. The black curves depict trial solutions, $S_\textnormal{trial}(q,\omega)$, that are consistent with our PIMC data [cf.~Eqs.~(\ref{eq:FS}) and (\ref{eq:mome})], whereas the colored curves are being discarded.
    Taken from Ref.~\cite{groth_prb_19} with the permission of the authors.
    }
    \label{fig:DSF_PRB}
\end{figure}

This is illustrated in Fig.~\ref{fig:DSF_PRB}, where we show the frequency dependence of $S(q,\omega)$ for a fixed wave number $q\approx2q_\textnormal{F}$ at $r_s=6$ and $\theta=1$. First and foremost, we note that the stochastic sampling of $G_\textnormal{trial}(q,\omega)$ still allows for nontrivial structures in $S(q,\omega)$, and even solutions with two peaks are possible. The colored curves correspond to those $S_\textnormal{trial}(q,\omega)$ that are not consistent with our PIMC data for $F(q,\tau)$ and $\braket{\omega^k}$, whereas the black curves are valid solutions, and are included in the calculation of the average for the final solution for $S(q,\omega)$. Moreover, we note that all black curves fall within a narrow band around their average, and exhibit a single broad peak centered around twice the plasma frequency. Therefore, the remaining degree of uncertainty is small, and we have achieved an accurate reconstruction of the DSF of the warm dense UEG.

\begin{figure}
    \centering
\includegraphics[width=0.5\textwidth]{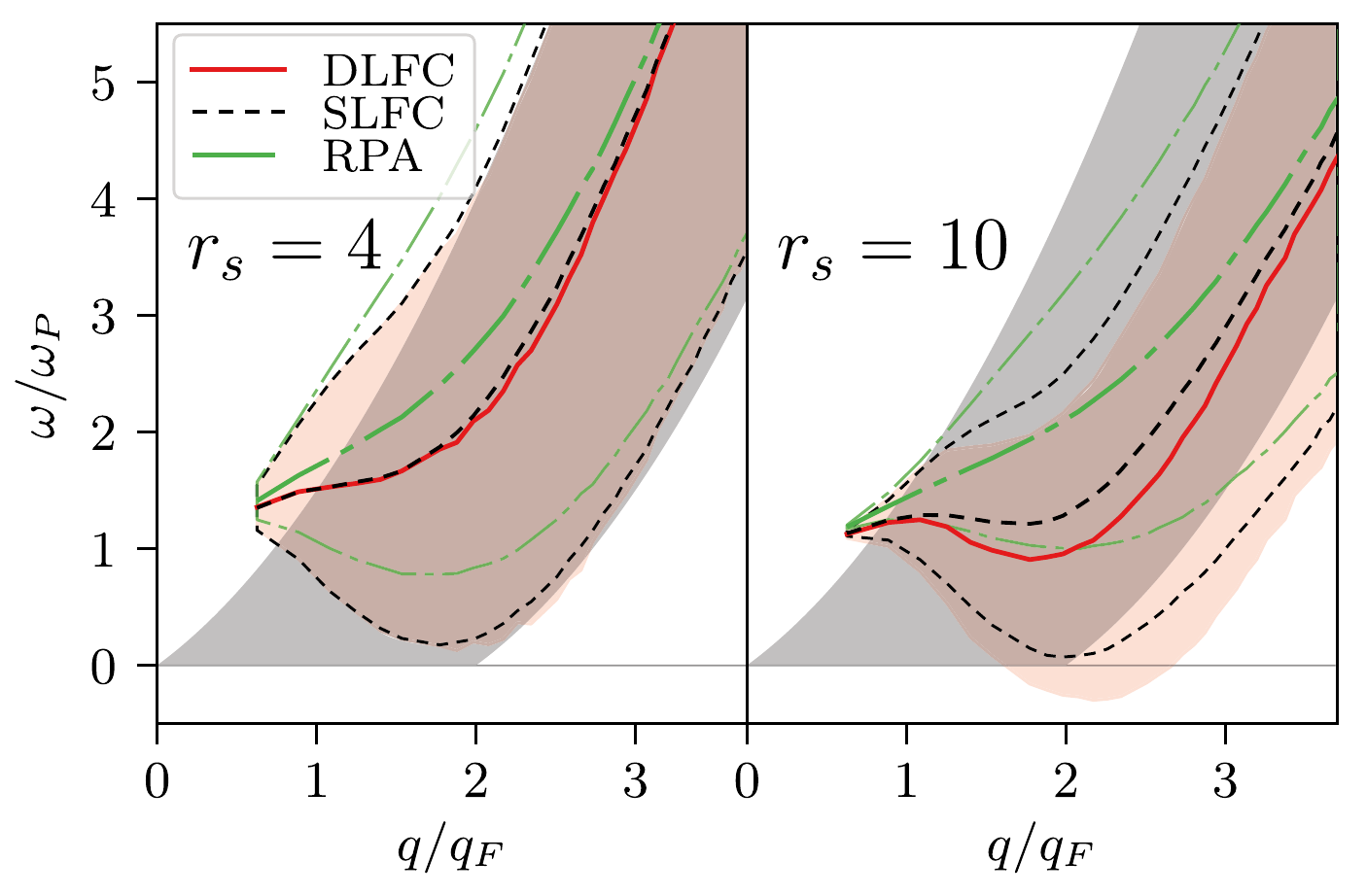}
    \caption{Dispersion relation of the unpolarized UEG at the Fermi temperature, $\Theta=1$, for $r_s=4$ (left) and $r_s=10$ (right). The solid red, black dashed and solid green curves depict the position of the maximum in $S(q,\omega)$ obtained from the full solution of our stochastic sampling procedure (dynamic LFC, DLFC), the \textit{exact} static approximation, Eq.~(\ref{eq:define_LFC_static})  (static LFC, SLFC), and the usual random phase approximation (RPA), respectively. The corresponding shaded areas illustrate the full width at half maximum, and the grey areas enclosed by two parabolas the zero temperature electron--hole pair continuum~\cite{giuliani2005quantum}.
    Taken from Ref.~\cite{dornheim_prl_18} with the permission of the authors.
    }
    \label{fig:DSF_PRL}
\end{figure}

These new \textit{ab initio} results for the dynamics of correlated electrons have opened up numerous new avenues for future research projects.
Most importantly, we mention that the detailed investigation of $S(q,\omega)$ is interesting in its own right and might, potentially, lead to the discovery of hitherto unobserved physical effects. This is illustrated in Fig.~\ref{fig:DSF_PRL}, where we show the dispersion relation of $S(q,\omega)$ at the Fermi temperature for two different values of the density parameter $r_s$. Let us first focus on the red and green curves (shaded areas), which depict the peak positions (full width at half maximum) of the \textit{ab initio} solution for $S(q,\omega)$ computed from the stochastic sampling of the dynamic LFC (DLFC) and the random phase approximation (RPA). At metallic density ($r_s=4$, left panel), the exact DSF exhibits a significant red-shift and correlation induced broadening as compared to the mean-field solution, which are particularly pronounced for intermediate wave numbers. On the other hand, all curves converge for small and large $q$, as expected. The right panel shows the same information for stronger coupling strength, $r_s=10$. Remarkably, we find a pronounced \textit{negative dispersion $\omega(q)$} in the DLFC curve around $q\approx1.8q_\textnormal{F}$, which is not captured by RPA at all. This feature had previously been reported by Takada and Yasuhara~\cite{Takada_PRL_2012,Takada_PRB_2016} based on approximate results at zero temperature, and might indicate an incipient excitonic mode, which emerges in the electron liquid regime. A more detailed investigation of this effect, which includes a "phase diagram" of its appearance regarding $r_s$ and $\theta$, and a prediction of experimental conditions, where it can be measured, is currently in progress. 
As a second important application, we mention the interpretation of XRTS experiments~\cite{redmer_glenzer_2009}, where the DSF of the UEG is used to describe the free electronic part~\cite{Kraus_2018}.

The third application of the new data for the dynamics of the warm dense UEG is their potential utility as input for other simulation methods. For example, the dynamic LFC is directly related to the exchange--correlation kernel of TD-DFT~\cite{baczewski_prl_16} via
\begin{eqnarray}
K_\textnormal{xc}(q,\omega) = - \tilde v_q G(q,\omega) \quad ,
\label{eq:kxc-g}
\end{eqnarray}
which gives rise to the intriguing possibility to systematically go beyond the nearly ubiquitous adiabatic approximation for the XC-functional. In addition, such information can directly be used to further improve QHD simulations where a similar relation as (\ref{eq:kxc-g}) has been derived in Ref.~\cite{zhandos_pop18}, cf. Sec.~\ref{ss:averaging}. 

\begin{figure*}
    \centering
\includegraphics[width=0.75\textwidth]{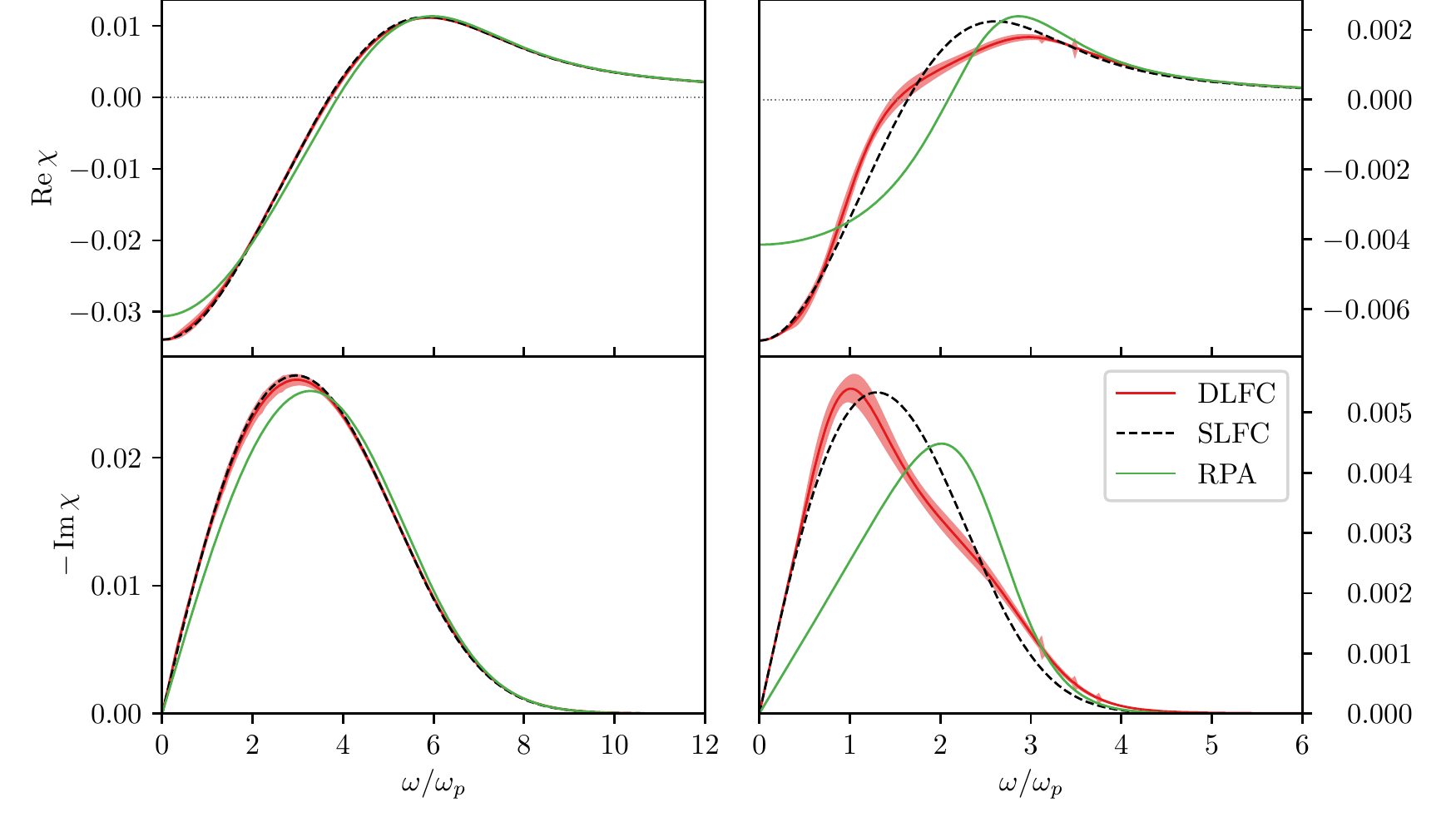}
    \caption{\textit{Ab initio} PIMC results for the dynamic density response function computed via Eq.~(\ref{eq:define_LFC}) for electrons under WDM conditions: $\Theta=1, k=1.88 k_F$. Top: real part, bottom: Imaginary part. Left (right) column: $r_s=2$ ($r_s=10$). The RPA result is compared to the full PIMC data (DLFC) and the static approximation (SLFC), Eq.~(\ref{eq:define_LFC_static}).}
    \label{fig:chi-dyn_paul}
\end{figure*}

The fourth application of our \textit{ab initio} dynamic results is the computation of additional material properties, such as the stopping power~\cite{Nagy1985,Stopping_Power_2017,Cayzac2017}, the dynamic conductivity~\cite{Veysman_PRE_2016,Desjarlais_PRE_2017},
 the dynamic dielectric function $\epsilon(q,\omega)$ and the density response function $\chi(q,\omega)$. As an example we show preliminary results for the dynamic density response function, Eq.~(\ref{eq:define_LFC}) in 
Fig.~\ref{fig:chi-dyn_paul}. Again we clearly see the effect of correlations, by comparing the dynamic results (red) to the RPA (green). While, for $r_s=2$ the effect is relatively small and mainly seen in a redshift of the imaginary part, at $r_s=10$, the RPA completely fails to describe the density response.

A fifth important application of the \textit{ab initio} data for $S(q,\omega)$ is that they unambiguously allow us to benchmark previous approximations~\cite{nozieres2018theory,kwong_prl_00,Kas_PRL_2017,Holas1987}, which are commonly used for WDM research. For example, Dornheim~\textit{et al.}~\cite{dornheim_prl_18} have reported that RPA exhibits significant inaccuracies even at relatively moderate coupling, $r_s=2$ and $\theta=1$, where electronic correlation effects had often been assumed to play a minor role. This has potentially important consequences for the interpretation of WDM experiments, as the determination of plasma parameters, such as the electronic temperature and the degree of ionization, is sensitive to the exact dispersion relation of the DSF of free electrons~\cite{Kraus_2018}. Moreover, it has allowed us to introduce a significantly more accurate, yet computationally equally cheap alternative to the RPA. More specifically, we replace in Eq.~(\ref{eq:define_LFC}) the dynamic LFC 
by its exact static limit, 
\begin{eqnarray}\label{eq:define_LFC_static}
\chi^{\rm SLFC}({q},\omega) = \frac{ \chi_0({q},\omega) }{ 1 - \tilde v_q\left[1-G({q},0)\right]\chi_0({q},\omega)} \; ,
\end{eqnarray}
that is conveniently available as a neural-net representation~\cite{dornheim_jcp_19-nn}, see Sec.~\ref{sec:LFC}. The corresponding results for the dispersion relation of $S(q,\omega)$ computed within this \textit{exact static approximation} (as opposed to static dielectric theories like STLS~\cite{sjostrom_dufty_2013,Tanaka_1985,Ichimaru_1987}, where the results for $G(q,0)$ are approximate and systematically biased) are shown as the dashed black lines in Fig.~\ref{fig:DSF_PRL}. At warm dense matter conditions ($r_s=4$ and $\theta=1$, left panel), the SLFC curve is basically indistinguishable from the exact data both with respect to peak position and shape over the entire $q$-range. Upon approaching the electron liquid regime ($r_s=10$, right panel), there do appear small yet significant deviations between the two curves, although the SLFC still captures both, the broadening and the negative dispersion. The same behavior is seen in the dynamic density response function, cf. Fig.~\ref{fig:chi-dyn_paul}. Thus we conclude that the \textit{exact static approximation} constitutes a distinct improvement over the RPA everywhere, without any additional effort.

Finally, an ambitious follow-up project regarding the \textit{ab initio} calculation of dynamic properties is the extension of our simulations to real WDM systems, i.e., going beyond the UEG model and to also include ions. Although computationally challenging, this would allow for the first \textit{exact} theoretical results for the dynamics of warm dense matter. More specifically, the combination of PIMC and the subsequent analytic continuation does not require any arbitrary external input, such as  the XC-functional in DFT, or the Chihara decomposition~\cite{Chihara_1987}, which presupposes a potentially unrealistic distinction between bound and free electrons~\cite{baczewski_prl_16}.

\section{Finite temperature DFT results}\label{s:dft}
\subsection{Kohn-Sham-Mermin DFT}\label{ss:dft-mermin}

In this section we explore the effect of the finite temperature exchange correlation functionals that were obtained by QMC simulations [cf. Sec.~\ref{sec:static}] in DFT simulations of dense plasmas. We present results for the equation of state (EOS) of dense hydrogen and carbon in Figs.~\ref{fig:dft_h2_electronic} and \ref{fig:dft_c_total_electronic}. 
\par
The finite temperature DFT-MD method combines the quantum treatment of the fast moving electrons with the classical description of the slow ion dynamics \cite{DFT_MD}. For the electrons, finite temperature DFT developed by Mermin \cite{mermin_1965} for the Kohn-Sham scheme~\cite{kohn_sham} is applied, which minimizes the grand potential,  $\Omega=E-TS-\mu N$. Here $E$ is the total energy, $T$ is the electron temperature, $S$ is the entropy, $\mu$ is the chemical potential, and $N$ is the number of electrons. For simplicity, $E$ and $S$ are expressed in spin-averaged form as 
\begin{equation}
  \begin{aligned}
E=&-\sum_{i=1}^{\infty}f^{\rm eq}(\epsilon_i)\bra{\psi_i}\nabla^2\ket{\psi_i}+E^H[n]+E^{xc}[n]\\
&+\int d\textbf{r}\,V^{ei}(\textbf{r})n(\textbf{r})
  \end{aligned}
\end{equation}
 and \begin{equation}
     S=-2\sum_{i=1}^{\infty}\{f^{\rm eq}(\epsilon_i)\text{ln}f^{\rm eq}(\epsilon_i)-[1-f^{\rm eq}(\epsilon_i)]\text{ln}[1-f^{\rm eq}(\epsilon_i)]\},
 \end{equation} 
where $i$ is the index of the energy eigenvalues, $E^H$ is the Hartree energy, $E^{xc}$ is the exchange-correlation energy, $V^{ei}$ is the ionic potential experienced by the electrons, and
$n(\textbf{r})$ is the charge density of the electrons. Further, 
\begin{align}
f^{\rm eq}_i = f^{\rm eq}(\epsilon_i)=\frac{1}{\text{exp}[\beta(\epsilon_i-\mu)]+1} \,,\;\; \beta=\frac{1}{k_BT}\,,   \label{eq:fermi-function}
\end{align}
 represents the Fermi-Dirac equilibrium distribution. The energies $\epsilon_i$, the wave functions $\psi_i$, the chemical potential $\mu$, and the charge density $n(\textbf{r})$ are self-consistently determined from the variational Kohn-Sham equation 
\begin{equation}
    \left[-\frac{1}{2}\nabla^2+U^{\rm H}[n]+V^{xc}[n]+V^{ei}(\textbf{r})\right]\psi_i(\textbf{r})=\epsilon_i\psi_i(\textbf{r})\,,
\end{equation}
with
\begin{equation}
    n(\textbf{r})=2\sum_{i=1}^{\infty}f^{\rm eq}(\epsilon_i)\lvert \psi_i(\textbf{r})\rvert ^2,
\end{equation}
and $\mu$ is determined by the charge conservation equation
\begin{equation}
    N=2\sum_{i=1}^{\infty}f^{\rm eq}(\epsilon_i)\,,
\end{equation}
where the orthonormality of the orbitals has been assumed.
When the Kohn-Sham-Mermin equations are solved self-consistently, the forces acting on each ion can be determined by the Hellman-Feynman theorem or its finite-temperature generalization. Then classical Newton's equations are solved to compute the dynamics of the  ions.
\par
In the finite temperature DFT (FT-DFT) method, the many-body effects of the electrons beyond the Hartree mean field are accounted for by
the exchange-correlation functional $E^{xc}$. In particular, if the exact functional
would be used, one could reproduce the exact solution of the original many-body
problem of interest. For practical applications, however, this term has to be approximated. Previous exchange-correlation functionals were limited to the case of zero temperature such as the expressions due to Perdew-Wang (PW)~\cite{PW92} and  Perdew-Burke-Ernzerhof (PBE)~\cite{PBE}. The latter are adequate for many condensed matter applications, but they become problematic when it comes to WDM. In this case finite temperature and entropic effects in the exchange-correlation functional are becoming important \cite{karasiev_2016,ramakrishna2020influence}, and, instead of $E^{xc}$, an accurate \textit{exchange-correlation free energy}, $F^{xc}$, has to be used. In this section, we quantitatively examine the importance of the related finite temperature effects.

\subsection{Equation of state of warm dense hydrogen}\label{ss:dft-hydrogen}
\begin{figure}
    \centering
\includegraphics[width=0.5\textwidth]{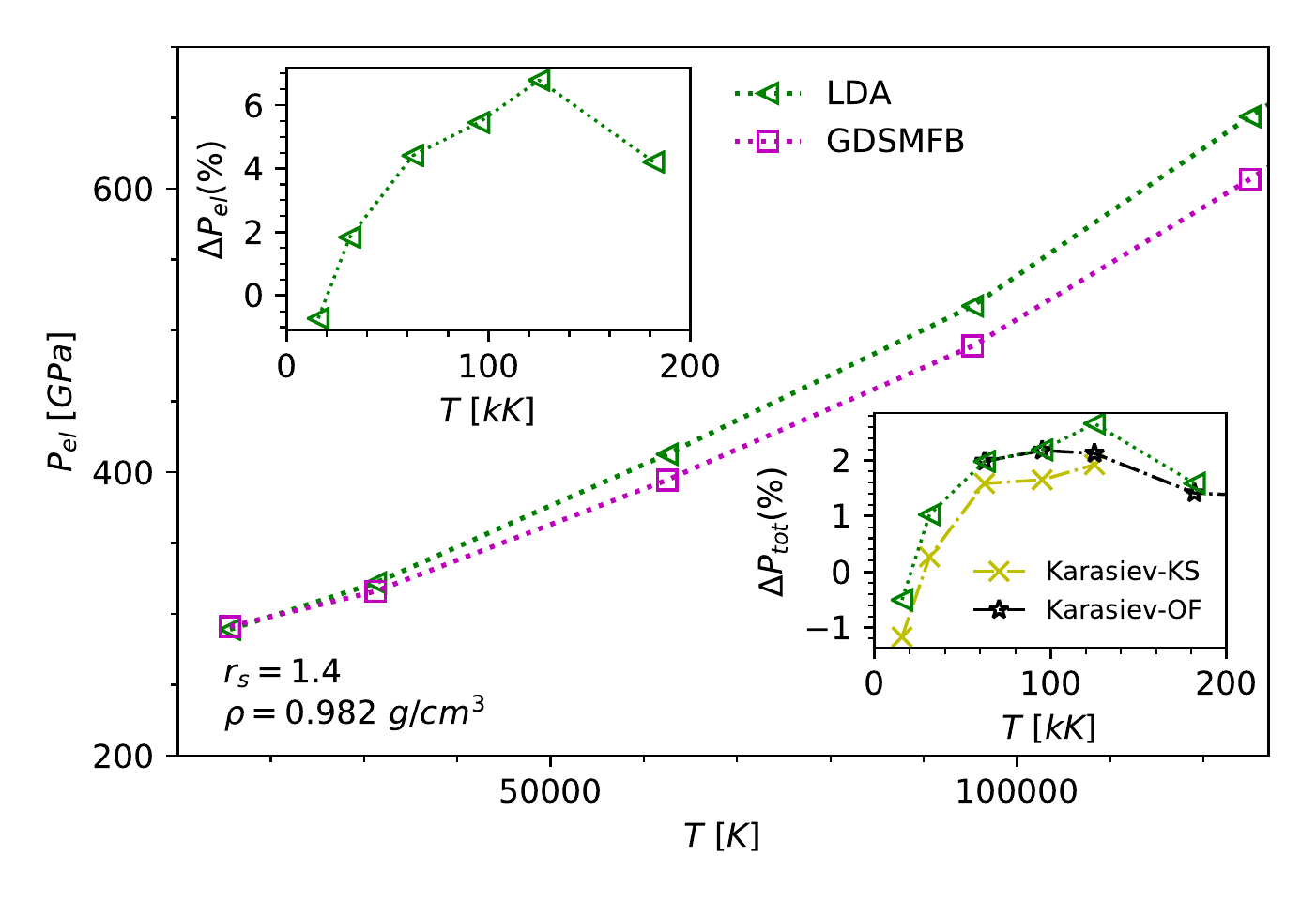}
\includegraphics[width=0.5\textwidth]{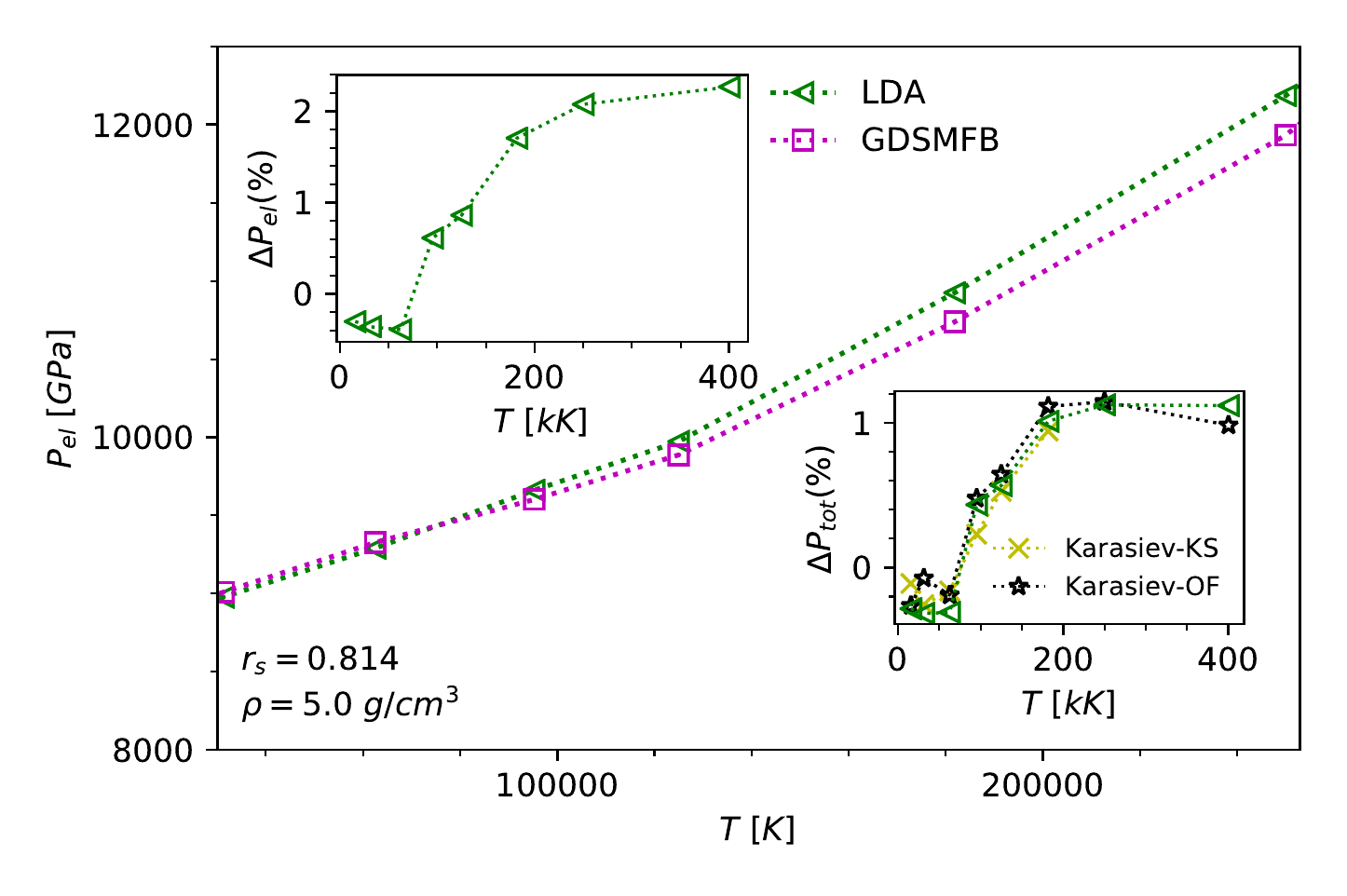} 
    \caption{Electronic part of the pressure of dense hydrogen using zero-temperature (LDA) and finite temperature (GDSMFB, Ref.~\cite{groth_prl17}) functionals. Left insets show the relative pressure difference,  $(p_{el}^{\rm LDA}-p_{el}^{\rm GDSMFB})/p_{el}^{\rm LDA}$. The insets on the bottom right show the relative difference in total pressure for GDSMFB with respect to LDA and also by Karasiev \textit{et al.} using Kohn-Sham (KS) and orbital-free (OF) DFT ~\cite{karasiev_2016}.} 
    \label{fig:dft_h2_electronic} 
\end{figure}   
The DFT-MD simulations for \textit{dense hydrogen} have been performed using the CP2K code~\cite{doi:10.1002/wcms.1159}. The Gaussian plane waves method is used to solve the Kohn-Sham equations with Gaussians as the basis set and  additional plane waves as the auxiliary basis of the form $\psi(\vec{r})=R_{i}(r)Y_{l_{i},m_{i}}(\theta,\phi)$, with $R_i(r)$ denoting the radial part and $Y_{l,m}$ denoting the angular part. Goedecker-Teter-Hutter pseudopotentials (GTH) of LDA (Pade) form are used for approximating the potential due to the usage of the LDA form of the xc-functional as the reference functional to compare with the parametrized LDA form incorporating finite temperatures, hereafter referred to as  GDSMFB~\cite{groth_prl17,PhysRevB.23.5048}. The GDSMFB functional is accessed in the CP2K code using the library of exchange-correlation functionals (LIBXC) commonly supported by DFT codes~\cite{MARQUES20122272,LEHTOLA20181}.  
\par 
Due to the huge computational cost at high densities and the large temperatures considered in these cases, we choose 
32
atoms in an hexagonal supercell for the simulations. On the same note, the sampling is performed only at the $\Gamma$-point. The system size and the sampling of \emph{k}-points can improve the convergence of the EOS, especially near the liquid-liquid phase transition (LLPT) and in the low density limit~\cite{PhysRevB.82.195107}. 
We observe finite size effects resulting in a lower electronic pressure, at low temperatures, compared to the orbital-free MD results of Wang \textit{et al.}~\cite{doi:10.1063/1.4821839}.  
\par 
At the low-temperature limit, especially near the LLPT ($1000-2000$) K, the pressure obtained using DFT-MD for these densities is dependent on a suite (Jacob's ladder, non-local, dispersion $\ldots$) of xc-functionals ignoring other parameters such as system size and \emph{k}-point sampling~\cite{knudson2017high,PhysRevB.89.184106}. The first order phase transition (LLPT) is well characterized using  $(\frac{\partial P}{\partial \rho})|_{T} = 0 $ in the EOS visibly more prominent in the QMC results compared to DFT-MD which requires more sampling near the transition region~\cite{morales2010evidence,PhysRevLett.120.025701}.  The synthesis of metallic hydrogen is among the current key topics  
of high-pressure physics, and  significant progress has been made over the last decade in the prediction of the transition using QMC and higher rungs of xc-functionals 
~\cite{celliers2018insulator,dias2017observation,knudson_direct_2015,pierleoni2016liquid}. The inclusion of the finite-temperature component to the LDA can be simply ignored for characterizing the phase transition at temperatures (1000-5000) K, as this corresponds to $\Theta<0.01$ and, instead, we focus on the improvement in the EOS results across a gamut of higher temperatures accessible using Kohn-Sham and orbital-free DFT.    
\par    
The variation of electronic pressure with respect to temperature at two different densities for dense hydrogen is shown in Fig. \ref{fig:dft_h2_electronic}. At $r_{s}=1.4,$ the total pressure with the finite-temperature xc-functionals differs by less than $0.5\%$ compared to LDA, in the temperature range (5000-10000) K, and accurate reptation quantum Monte Carlo (CEIMC) results show a deviation of less than $2\%$~\cite{morales2010equation}. With increasing temperature, the pressure obtained using GDSMFB converges towards the path integral Monte Carlo results obtained by Hu \textit{et al.}~\cite{hu_militzer_PhysRevLett.104.235003, PhysRevB.84.224109}. The relative difference in electronic pressure due to finite-temperature xc effects is more prominent at lower densities, with a maximum of $6\%$, observed at $r_{s}=1.4$, for $\Theta=0.33-0.42$. The relative difference in total pressure is in good agreement with the Kohn-Sham DFT and orbital-free results obtained by Karasiev \textit{et al.}~\cite{karasiev_2016}.   
An analysis of finite temperature exchange-correlation effects on various optical and transport properties of deuterium was recently presented in Ref.~\cite{karasiev_PhysRevB.99.214110}, and an extensive topical investigation of hydrogen can be found in Ref.~\cite{ramakrishna2020influence}.

\subsection{Equation of state of warm dense carbon}\label{ss:dft-carbon}
The DFT-MD simulations for dense carbon are performed using the recently developed ext-FPMD method \cite{extFPMD} implemented in the \textsc{Quantum-Espresso} code \cite{QE}, which combines the analytical treatment of high-energy electrons as plane waves and the numerical treatment of the remaining electrons within Kohn-Sham-Mermin scheme. This ext-FPMD method thus elevates the temperature limit of previous DFT-MD simulations and can be coherently applied from cold materials to hot dense plasmas \cite{gao2016}. The interaction between the carbon ions and the electrons is described by an all-electron PAW potential \cite{paw}. 32 carbon atoms are included in our simulation, which amounts to 192 electrons in total, in the cubic simulation box with periodic boundary conditions for all three directions. A shifted $2\times2\times2$ K mesh grid is used for all the simulations \cite{MPmesh}. The ion temperature is controlled by an Andersen thermostat \cite{andersen_thermo}. A sufficiently large number of time steps are applied to ensure that the system has reached equilibrium before data collection starts using the last 5000 time steps. The electronic part of pressure is converged to within 1\% with respect to all parameters such as plane wave cutoff energy, K-mesh density, and finite size effects. 

\par

\begin{figure} 
    \centering
\includegraphics[width=0.54\textwidth]{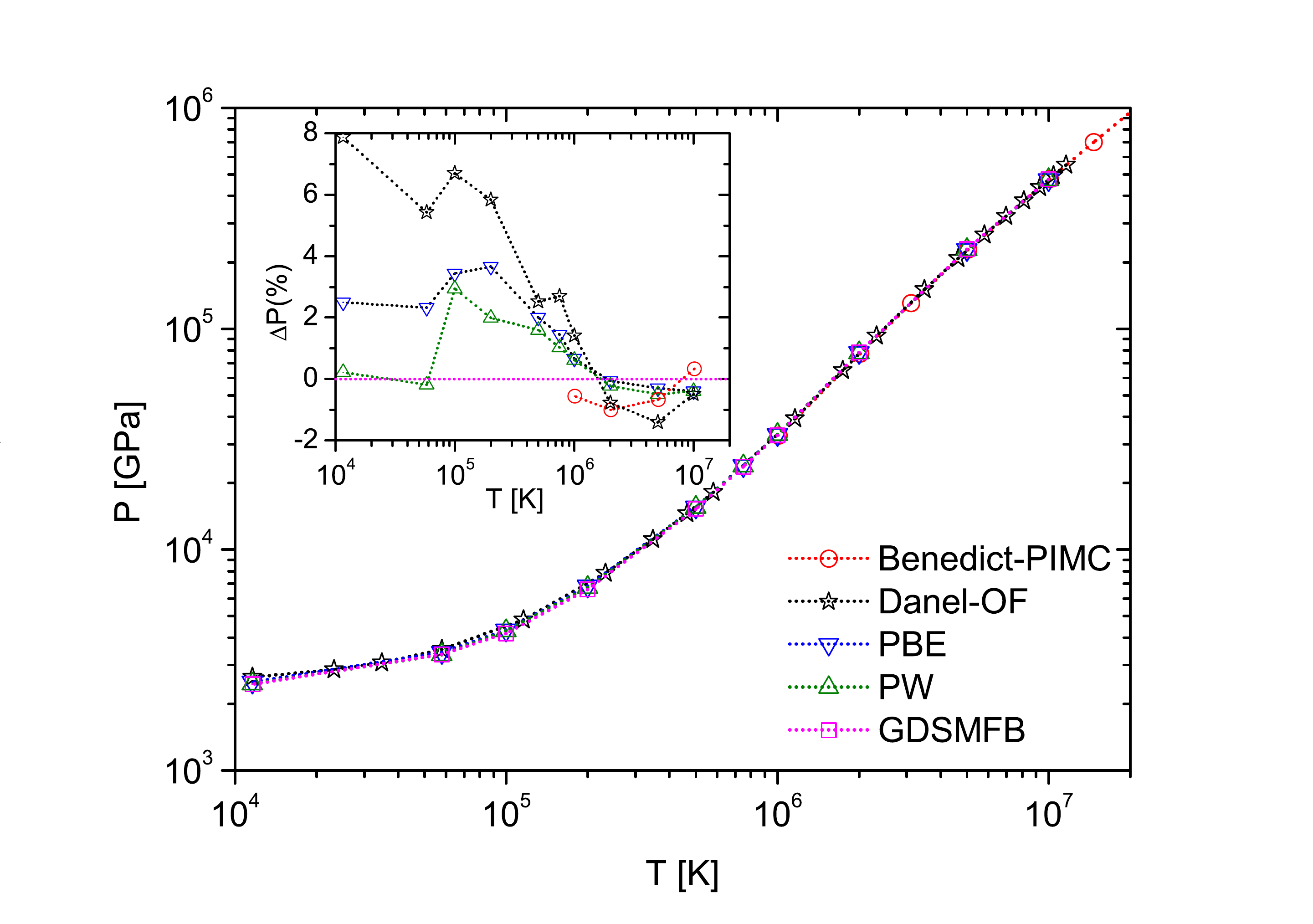}
\includegraphics[width=0.54\textwidth]{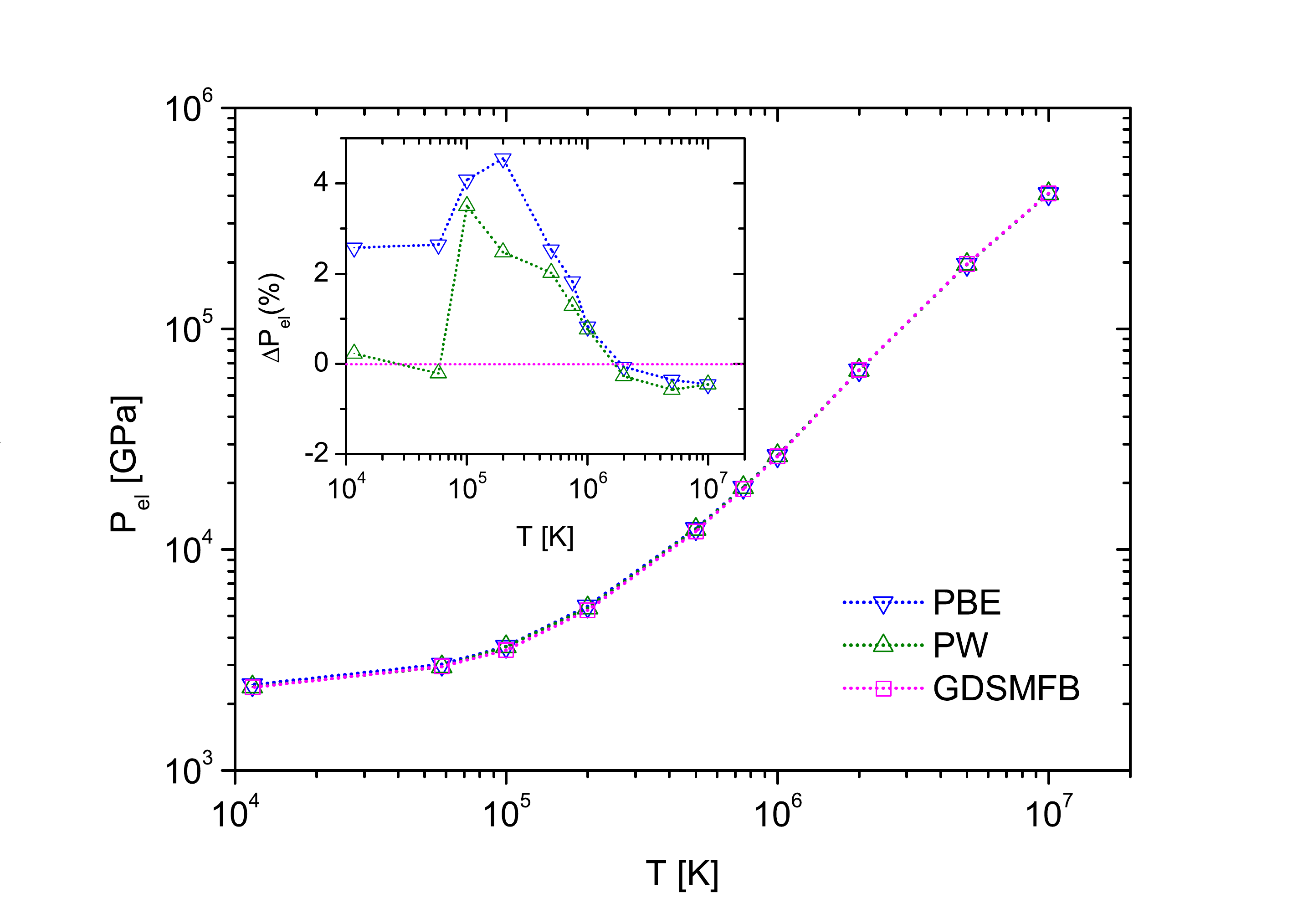}
    \caption{Equation of state of warm dense carbon for $\rho=10.0\, $gcm$^{-3}$ ($r_s=1.4755$), computed with various models. Top: total pressure, bottom: electronic part of pressure. Insets  show the relative difference of various models to our finite-temperature results (GDSMFB, Ref.~\cite{groth_prl17}). PW: zero-temperature LDA (PW-functional)~\cite{PW92}; PBE: zero-temperature PBE-functional~\cite{PBE}); Danel-OF: orbital-free MD results~\cite{danel2018}; Benedict-PIMC: restricted PIMC results~\cite{benedict2014}. Interpolation is applied to align the different data grid when necessary.} 
    \label{fig:dft_c_total_electronic}
\end{figure}

The variation of both, total pressure and electronic part of pressure of carbon, at a density of $\rho=10.0$ g/cm$^3$, corresponding to $r_s=1.4755$, is shown in Fig. \ref{fig:dft_c_total_electronic}. For the lowest temperatures, i.e. at 1eV and 5eV, we find that LDA~\cite{PW92} and the finite temperature GDSMFB results~\cite{groth_prl17} are close to each other. Both deviate from the PBE result~\cite{PBE}, which shows that the gradient correction of the exchange-correlation energy is more important than the finite-temperature effects, for low temperature, as expected. As the temperature rises, LDA and PBE results get closer to each other, however, both deviate from the GDSMFB result. This shows that, in this region,  finite-temperature effects play a more important role. The relative deviation for the electronic part of the pressure between zero-temperature exchange-correlation functionals and their finite-temperature counterpart reach a maximum of $\sim$4\% at 10$^5$ K for PW and $2\times10^5$ K for PBE, where $\Theta=0.374$ and $\Theta=0.749$, respectively. 

We note that a further increase of the temperature will eventually make the form of the exchange-correlation functionals less important, as the system approaches the hot classical plasma regime where many-body effects are less prominent. This is shown in Fig.~\ref{fig:dft_c_total_electronic} for the high-temperature region around 10$^7$ K. PIMC results \cite{benedict2014} are only available for high temperatures, and they are within 1\% to our GDSMFB results shown in this figure. The deviations are comparable to the data accuracy, due to the statistical errors. OFMD simulations~\cite{danel2018} struggle, in the low-temperature region, because they lack shell structure effects, and they are also found to be inaccurate in the high-temperature region, because they use the PBE exchange-correlation functional that does not account for finite-temperature effects. 
%

\section{WDM out of equilibrium}\label{s:nonequilibrium}
The response of warm dense matter to an external excitation and the subsequent thermalization are of prime importance for many applications. This includes laser excitation and ionization of warm dense matter but also compression experiments including phase transitions and the path to inertial confinement fusion.

Theoretical methods for WDM out of equilibrium are even more challenging than equilibrium applications that were discussed above. They include time-dependent DFT \cite{baczewski_prl_16,medvedev_prb15}, semiclassical kinetics \cite{rethfeld_cpp19}, quantum kinetic theory and nonequilibrium Green functions \cite{semkat_99_pre,semkat_00_jmp,vorberger_pre18, schluenzen_19_prl}, hydrodynamics \cite{bornath_cpp19,zhandos_pop18,PhysRevE.92.053112} or rate equations \cite{ohde_pop_95}. Among the problems that were studied are the equilibration of the electron distribution by electron-electron collisions \cite{semkat_99_pre,semkat_00_jmp},  non-thermal melting induced by fs x-ray pulses \cite{medvedev_prb15}, the
density response for nonequilibrium momentum distributions \cite{vorberger_pre18,gericke_prl11}, density evolution following short-pulse laser excitation \cite{rethfeld_cpp19}, collisional heating of quantum plasmas by a laser pulse \cite{haberland_01_pre,bonitz_99_cpp}, or
 ionization dynamics in a short laser pulse \cite{bornath_cpp19}. 
 
 In the reminder of this section we discuss the quantum hydrodynamics approach and its relation to DFT \cite{bonitz_pop_19} more in detail because the former is comparatively little discussed for WDM applications, even though it appears to be filling a gap in the arsenal of simulation techniques, what we discuss in Sec.~\ref{s:conclusion}.

\subsection{Dynamics of N quantum particles. Wave function and density operator}\label{ss:tdse}
We consider a non-relativistic  quantum system of $N$ electrons described by the spin-independent hamiltonian
\begin{eqnarray}
\hat H = \sum_{i=1}^N \left( -\frac{\hbar^2}{2m}\nabla^2_i + V(\textbf{r}_i) \right) + \frac{1}{2}\sum_{i\ne j} w_{ij}(\textbf{R}),
\label{eq:hamiltonian}
\end{eqnarray}
where $\textbf{R}=(\textbf{r}_1,\sigma_1;\textbf{r}_2,\sigma_2;\dots \textbf{r}_N,\sigma_N)$, $\textbf{r}_i$ are the particle coordinates and $\sigma_i$ their spin projections, and $V$ is an external potential, e.g. due to the plasma ions. 
Assuming first as pure state,  the dynamics of the system are governed by the N-particle Schr\"odinger equation
\begin{equation}
    i\hbar\frac{\partial \Psi(\textbf{R},t)}{\partial t} = \hat H \Psi(\textbf{R},t), \quad \Psi(\textbf{R},t_0) = \Psi_0(\textbf{R})\,,
    \label{eq:tdse}
\end{equation}
that is supplemented by an initial condition and the normalization $\sum_{\sigma_1\dots \sigma_N}\int d^{3N}R\, |\Psi(\textbf{R},t)|^2=N$.
For particles with spin $s$, there are $g_s=2s+1$ different spin projections, and each spin sum gives rise to a factor $g_s$  (in the following, we will not write the spin arguments and spin sums explicitly).

If the many-body system 
(\ref{eq:hamiltonian}) is coupled to the environment -- as is typically the case in plasmas that we concentrate on in this section -- a description in terms of wave functions and the Schr\"odinger equation (\ref{eq:tdse}) is no longer adequate. Instead, the system is described by an incoherent superposition of wave functions (``mixed state''). This can be taken into account, by replacing the $N$-particle wave function by the $N$-particle  density operator \cite{bonitz_qkt},
\begin{equation}
    \hat \rho(t) = \sum_a p_a |\Psi^a(t)\rangle\langle \Psi^a(t)|\,, \quad {\rm Tr} \, \hat \rho(t) = 1\,,
\label{eq:rho-def}
\end{equation}
where the sum runs over projection operators on all solutions of the Schr\"odinger equation (\ref{eq:tdse}), and $p_a$ are real probabilities, $0\le p_a\le 1$, with $\sum_a p_a =1$. Here we used a general representation-independent form of the quantum states. It is directly related to the wave functions if the  coordinate representation is being applied: $\langle \textbf{R}|\Psi^a(t)\rangle=\Psi^a(\textbf{R},t)$ [$\langle \textbf{R}|$ are eigenstates of the coordinate operator in N-particle Hilbert space].  The previous case of a pure state is naturally included in definition (\ref{eq:rho-def}) by setting $p_k=1$ and all $p_{a\ne k}=0$. The second relation (\ref{eq:rho-def}) is the normalization condition where the trace denotes the sum over the diagonal matrix elements of $\hat \rho$, see below.
The equation of motion of $\hat \rho$ follows from the Schr\"odinger equation (\ref{eq:tdse}) and is  the von Neumann equation supplemented by the initial condition,
\begin{eqnarray}
i\hbar \frac{\partial}{\partial t}\hat \rho &-& [\hat H,\hat \rho] = 0,
\label{eq:von-neumann}
\\
\hat \rho(t_0) &=& \sum_a p_a |\Psi^a_0\rangle\langle \Psi^a_0|\,.
\end{eqnarray}

The method of density operators is well established in quantum many-body and kinetic theory, and relevant representations are the coordinate representation, momentum and Wigner representation, e.g.  \cite{bonitz_qkt,bonitz_aip_12,khan_springer_14}. 
From the N-particle density operator all time-dependent properties of a quantum system can be obtained. However, in many cases simpler quantities are sufficient such as reduced $s$-particle density operators, including the single-particle density operator (which is related to the distribution function or Wigner function) \cite{bonitz_qkt}: %
\begin{equation}
\hat F_1(t) \equiv  N{\rm Tr}_{2\dots N} \hat \rho(t)\,, \qquad {\rm Tr}_1\hat F_1 = N\,.
\label{eq:f1-def}
\end{equation}
The equation of motion for $\hat F_1$ follows straightforwardly from Eq.~(\ref{eq:von-neumann}), e.g.  Refs.~\cite{bonitz_qkt, bonitz_pop_19}, and is given below, cf. Eq.~(\ref{eq:f1-hartree}).

\subsection{Time-dependent Kohn-Sham equations for electrons in WDM}\label{ss:tdse-plasma-dft}
We now derive the equation of motion for the time-dependent single-particle orbitals of $N$ interacting electrons. We follow the idea of DFT that the many-particle quantities are expressed in terms of single-particle quantities via a mean field description. Exchange and correlation effects are then taken into account \textit{a posteriori}, by adding the potential $V^{\rm xc}$. 

Considering an N-particle system in the grand canonical ensemble (specified by the inverse temperature $\beta$ and chemical potential $\mu$), the single-particle nonequilibrium density operator  has the form
\begin{eqnarray}
\hat F_1(t; \beta,\mu) 
&=& \sum_{i=0}^{\infty} f^{\rm eq}_i(\beta,\mu)|\phi_i(t)\rangle \langle \phi_i(t)|\,,
\label{eq:f1}
\end{eqnarray}
where the mean occupation numbers in equilibrium are  given by the Fermi function (\ref{eq:fermi-function}).
The equation for $\hat F_1$ in the mean field (Hartree) approximation has the form \cite{bonitz_qkt}, 
\begin{align}
    i\hbar \frac{\partial \hat F_1}{\partial t} &- 
    [\hat H_1 + \hat H_1^{\rm H},\hat F_1] = 0\,,
    \label{eq:f1-hartree}\\
    \hat H_1^{\rm H} &= \mbox{Tr}_2 \hat w_{12}\hat F_2\,.
    \label{eq:hartree-op}
\end{align}
Correlation effects would give rise to a collision integral on the r.h.s. of Eq.~(\ref{eq:f1-hartree}), for various approximations, see Ref.~\cite{bonitz_qkt}.

From Eq.~(\ref{eq:f1}) we obtain the density matrix by multiplying with coordinate eigenstates $\langle \textbf{r}'|$ and $|\textbf{r}''\rangle$:
\begin{eqnarray}
f(\textbf{r}',\textbf{r}'',t;\beta,\mu) = \sum_{i=0}^{\infty} f^{\rm eq}_i(\beta,\mu)\phi_i(\textbf{r}',t)\phi^*_i(\textbf{r}'',t)\,. 
\label{eq:f1matrix}
\end{eqnarray}
The single-particle wave functions are the so-called ``natural orbitals'', and in the mean field approximation, the $N$-particle wave function obeying Eq.~(\ref{eq:tdse}) is just their product.
Inserting the ansatz (\ref{eq:f1matrix}) into the coordinate representation of Eq.~(\ref{eq:f1-hartree}),
it is easy to verify that the latter  is solved when each orbital fulfills the following single-particle Schr\"odinger equation ($i=1\dots N$)
\begin{eqnarray}
i\hbar\frac{\partial}{\partial t}\phi_i(\textbf{r},t) &=& \left\{-\frac{\hbar^2}{2m}\nabla^2_{\textbf{r}} 
+ V + U^{\rm H}(\textbf{r},t)\right\}\phi_i(\textbf{r},t)\,,\quad
\label{eq:td-ks}\\
U^{\rm H}[n(\textbf{r},t)] &=& g_s\int d\textbf{r}_2\, w(\textbf{r}-\textbf{r}_2) n(\textbf{r}_2,t;\beta,\mu),
\label{eq:tdks-hartree}\\
n(\textbf{r},t;\beta,\mu) &=& \sum_{i=0}^\infty f^{\rm eq}_i(\beta,\mu) |\phi_i(\textbf{r},t)|^2\,,
\label{eq:density_finite-t}
\end{eqnarray}
where the Hartree mean field is the coordinate representation of the operator (\ref{eq:hartree-op}) and contains the densities of all occupied orbitals.
Equations (\ref{eq:td-ks}) and (\ref{eq:tdks-hartree}) are the time-dependent Hartree equations for weakly interacting fermions (interactions are taken into account only via the mean field $U^{\rm H}$). 

This result can be directly extended beyond the mean field approximation by replacing 
\begin{align}
    U^{\rm H}(\textbf{r},t) & \to 
    U^{\rm H}[n(\textbf{r},t)] + V^{\rm xc}[n(\textbf{r},\tilde t);\beta,\mu]\,,
    \label{eq:xc-potential}
\end{align}
and, as a consequence, Eqs.~(\ref{eq:td-ks}), (\ref{eq:tdks-hartree}) become the  time-dependent Kohn-Sham equations--the basic equations of time-dependent density functional theory (TD-DFT) \cite{runge-gross}. A particular strength of this theory is its solid theoretical foundation on the Runge-Gross theorem \cite{runge-gross} and the corresponding theorems for time-independent DFT \cite{hohenberg-kohn}. The basic statement is that a system of $N$ interacting fermions can be mapped exactly on a system of $N$ non-interacting particles with the same density $n(\textbf{r},t)$ where all interactions are lumped into an effective single-particle potential that is a direct generalization of the Hartree potential (\ref{eq:tdks-hartree}).

The first remarkable property of these equations is that, both, the mean field and the additional exchange-correlation potential do not explicitly depend on the individual orbital wave functions but only on the total density, so also the coordinate dependence is only implicit, via the functional $n(\textbf{r},t)$. 
However, the exact functional $V^{\rm xc}$ does not only depend on the current density, $n(\textbf{r},t)$, but, in general the dependence is also on the density profile at earlier times, $n(\textbf{r},\tilde t), \; 0\le\tilde t\le t $. At the same time, most current implementations neglect this ``memory'' effect and use an adiabatic approximation (e.g. adiabatic LDA, ALDA), $\tilde t \to t$ which leads to systematic errors.
In Eq.~(\ref{eq:xc-potential}) we also indicated that, at finite temperature, $V^{\rm xc}$ carries a temperature dependence which was discussed in detail before, see Sec.~\ref{s:qmc}.

\subsection{Microscopic quantum hydrodynamic equations for dense plasmas}\label{ss:plasma-mqhd}
Following Ref.~\cite{bonitz_pop_19}, we now derive the  \textit{microscopic QHD (MQHD) equations}, starting from the time-dependent Hartree equations (\ref{eq:td-ks}) and (\ref{eq:tdks-hartree}).  To this end we simply convert each orbital solution, $\phi_i(\textbf{r},t)=A_i(\textbf{r},t)e^{\frac{i}{\hbar}S_i(\textbf{r},t)}$, into an individual pair of amplitude and phase equations  \cite{manfredi_fields_05,bonitz_pop_19}, for $i=1\dots N$,
\begin{eqnarray}
\frac{\partial n_i}{\partial t} + \nabla \cdot(\textbf{v}_in_i) &=& 0,
\label{eq:density-balance-fermions_i}
\\
\frac{\partial \textbf{p}_i}{\partial t} + (\textbf{v}_i\cdot \nabla) \textbf p_i &=& -\nabla(U_{\rm tot}[n]+V^{\rm xc}[n]+Q_i)\,,
\label{eq:momentum-balance-fermions_i}
\\
Q_i(\textbf{r}) &=& - \frac{\hbar^2}{2m}\frac{\nabla^2 \sqrt{n_i(\textbf{r})}}{\sqrt{n_i(\textbf{r})}}\,,
\label{eq:qpot-fermions_i}
\end{eqnarray}
where $n_i=A_i^2$ and $\textbf{p}_i=\nabla S_i$, and we introduced a short notation for the total potential energy, $U_{\rm tot}=V+U^{\rm H}$. This system  of MQHD equations 
is fully equivalent to TD-DFT and, for $V^{\rm xc}\to 0$, it exactly coincides with the time-dependent nonlinear Hartree (quantum Vlasov) equations \cite{bonitz_qkt,rukhadse_ufn_99}.
Moreover, it was shown in Ref.~\cite{bonitz_pop_19} that this approximation, in linear response,  is exactly equivalent to the random phase approximation (RPA or linearized quantum Vlasov equation). In particular, the linearized MQHD equations then yield the correct plasmon spectrum and the correct screening of a test charge -- in contrast to the standard QHD (see below). 

\subsection{Derivation of the QHD equations from MQHD}\label{ss:averaging}
To convert these microscopic equations into a single pair of density and momentum equations (QHD), a suitable averaging over the orbitals is necessary which we denote by a ``bar'':
\begin{eqnarray}
\overline{n}(\textbf{r},t) &=& \frac{1}{N}\sum_{i=1}^\infty f^{\rm eq}_i \, n_i(\textbf{r},t)\,,
\\
\overline{\textbf{p}}(\textbf{r},t) &=& \frac{1}{N}\sum_{i=1}^\infty f^{\rm eq}_i \, \textbf{p}_i(\textbf{r},t)\,,
\label{eq:mean-p}
\\
\overline {Q}(\textbf{r},t) &=& - \frac{\hbar^2}{2mN}\sum_{i=1}^\infty f^{\rm eq}_i \, \frac{\nabla^2 \sqrt{n_i(\textbf{r})}}{\sqrt{n_i(\textbf{r})}}\,,
\label{eq:mean-bohm-pot}
\end{eqnarray}
where the orbitals are weighted by the Fermi function. Here $\overline{n}$ is interpreted as the mean orbital probability density.
In Ref.~\cite{manfredi_prb_01} the authors assumed that all orbital amplitudes  are equal whereas, in Ref.~\cite{manfredi_fields_05}, it was assumed that one can substitute %
\begin{align}
\sum_{i=1}^\infty f^{\rm eq}_i \, \frac{\nabla^2 \sqrt{n_i(\textbf{r})}}{\sqrt{n_i(\textbf{r})}} \longrightarrow  \frac{\nabla^2 \sqrt{\overline{n}(\textbf{r})}}{\sqrt{\overline{n}(\textbf{r})}}\,. 
\label{eq:manfredie-assumption}
\end{align}
Finally, in Ref.~\cite{bonitz_pop_19} it was demonstrated how the QHD equations can be derived without uncontrolled assumptions, and here we briefly recall that approach. 

In order to take the orbital average of the MQHD equations (\ref{eq:density-balance-fermions_i}, \ref{eq:momentum-balance-fermions_i}, \ref{eq:qpot-fermions_i}), we express each of the orbital quantities in terms of their averages and fluctuations:
\begin{align}
n_i &=\overline n+\delta n_i\,,    
\nonumber\\
\textbf{p}_i &=
\overline{\textbf{p}}+\delta \textbf{p}_i\,,  \nonumber\\
A_i &=\overline A+\delta A_i\,,    
\nonumber
\end{align}
and take into account that the average of products of two orbital quantities is given by 
\begin{align}
\overline{a_i b_i} & =\overline{a}\cdot\overline{b} + \overline{\delta a_i \cdot \delta b_i}\,,
\label{eq:ab-correlation}
\end{align}
where, in addition to the product of averages, there appears a correlation function. Note that the averaging of the Bohm potentials, Eq.~(\ref{eq:mean-bohm-pot}), has to be done with care because the orbital depending densities enter at two places and we have to apply relation (\ref{eq:ab-correlation}),
with the result
\begin{align}
\overline{Q} &=Q_1[\overline{n}]+Q^\Delta\,.
\label{eq:q-fluctuation}
\end{align}
Here the first term is just the Bohm potential from the single-particle case with the density replaced by the mean density, $n_i \to \overline{n}$ (this is what was used in Refs.~\cite{manfredi_prb_01,manfredi_fields_05}), corresponding to Eq.~(\ref{eq:manfredie-assumption}), and the second term is the deviation which is presented below, in Eq.~(\ref{eq:delta-qpot-fermions_average}).
With this we can perform the averaging of the MQHD equations (\ref{eq:density-balance-fermions_i}, \ref{eq:momentum-balance-fermions_i}, \ref{eq:qpot-fermions_i}), and obtain the QHD equations that contain three correlation functions that we denote by $J_{np}, J_{pp}$ and $Q^\Delta$,
\begin{eqnarray}
&& \frac{\partial \overline{n}}{\partial t} + 
\nabla\cdot (\overline{\textbf{v}}\cdot\overline{n}) = J_{np}\,,
\label{eq:density-balance-fermions_average}
\\
&&\frac{\partial \overline{\textbf{p}}}{\partial t} + 
(\overline{\textbf{v}}\cdot\nabla)
\overline{\textbf p} = -\nabla \left(U_{\rm tot}+Q_1[\overline{n}] + Q^\Delta \right) + J_{pp}\,, \quad
\label{eq:momentum-balance-fermions_average}
\\
&& J_{np} = -\frac{1}{m}\nabla\cdot(\overline{\delta \textbf{p}_i  \delta n_i})\,, \quad
J_{pp} = - 
\overline{(\delta \textbf{v}_i\cdot
\nabla)\delta \textbf{p}_i}\,,\quad 
\\
&& Q^{\Delta} \approx  \frac{\hbar^2}{2m \overline{n}}\overline{\delta A_i \cdot\nabla^2\delta A_i} + O\left( \left(\frac{\delta A_i}{\overline{A}} \right)^2\right)\,.
\label{eq:delta-qpot-fermions_average}
\end{eqnarray}
The function $J_{pp}$ contains the correlations of the fluctuations of the momentum field.

Another formulation of the QHD equations, that is closer to classical hydrodynamics, is obtained if, instead of the mean orbital density $\overline{n}$ and the mean momentum $\overline{\vec p}$, we consider the density $n(\vec r,t)$ and the current density $\vec j(\vec r,t)$, defined in terms of the orbital quantities $n_i$ and $\vec j_i=n_i\,\vec v_i$, as~\cite{manfredi_fields_05, palade2018prb}
\begin{align}
    n(\vec r,t)&=2\sum_i f_i \, n_i(\vec r,t),\\
    \vec j(\vec r,t)&=2\sum_i f_i \, \vec j_i(\vec r,t),
\end{align}
where a factor $2$ was included to account for the two electron spin projections. Using the definitions for $n$ and $\vec j$, we can further define a mean velocity field, $\vec u(\vec r,t)=\vec j(\vec r,t)/n(\vec r,t)$ which differs from the mean velocity $\overline{\vec v}=\overline{\vec p}/m$ that was defined above, cf. Eq.~(\ref{eq:mean-p}).

The dynamical equations for $n$ and $\vec j$ follow from Eqs.~\eqref{eq:density-balance-fermions_i},~\eqref{eq:momentum-balance-fermions_i} and~\eqref{eq:qpot-fermions_i}. They read as (see also Ref.~\cite{palade2018prb})
\begin{align}
    \frac{\partial n}{\partial t}+\nabla\cdot \vec j&=0,
    \label{eq:continuity}\\
    \frac{\partial \vec j}{\partial t}+\nabla\cdot \left(\frac{\vec j \otimes \vec j}{n}\right)&=-\frac{n}{m}\nabla U_{\rm tot}[n]-\frac{1}{m}\nabla\cdot \vec \Pi\label{eq:momentum},
\end{align}
where $\vec \Pi(\vec r,t)$ appears as a ``pressure'' tensor,
\begin{align}
    \vec \Pi(\vec r,t) = 2\sum_i f_i \,n_i \bigg[ m  (\vec v_i-\vec u)\otimes (\vec v_i-\vec u)   \nonumber \\
    \left. -\frac{\hbar^2}{4m}\nabla\otimes\nabla \ln n_i \right].
\end{align}
The first contribution arises from fluctuations of the orbital velocity fields while the second term is due to the microscopic Bohm potential, i.e. due to the curvature of the orbital amplitudes. Note that the correlation function $J_{np}$ appearing in Eq.~\eqref{eq:density-balance-fermions_average} is contained in the definition of the current density and does not appear explicitly in Eq.~\eqref{eq:continuity}. Analogous equations have been discussed in Ref.~\cite{palade2018prb}, see also Refs.~\cite{tokatly2009, ciraci2017prb}.

\subsection{Plasma oscillations in MQHD and QHD}\label{ss:plasma_oscillations}
An important test for the QHD and MQHD models is the result for electron plasma oscillations (Langmuir waves) in the limit of weak external field (linear response). Here we consider the simplest case of a spatially homogeneous weakly non-ideal electron gas (interactions are included only via the Hartree mean field whereas exchange-correlation effects are neglected) at zero temperature, i.e. the statistical weights $f^{\rm eq}_i$ reduce to unity, for $\epsilon_i \le E_F$, and zero otherwise.

Considering first the MQHD equations, Eqs.~(\ref{eq:density-balance-fermions_i}, \ref{eq:momentum-balance-fermions_i}), and (\ref{eq:qpot-fermions_i}),  we apply a harmonic monochromatic excitation, $\phi_1(\textbf{r},t)\sim e^{-i\omega t+i \textbf{q}\cdot \textbf{r}}$, and linearize $n_i(t)$  and $\vec p_i(t)$ around the unperturbed solution. Finally, the density response is computed via orbital averaging and Fourier transformation, with the result given by Eq.~(\ref{eq:mqhd-dispersion}). Second, we consider the QHD equations~\eqref{eq:continuity}, \eqref{eq:momentum},
which already contain the orbital averaging. Here, in order to make further progress, an approximation for $\vec \Pi$ is required. If we approximate the tensor by the Fermi pressure and the Bohm term [with the density $n(\vec r,t)$],
\begin{align}\label{eq:approxpress}
    \vec \Pi(\vec r,t)\approx p_F[n(\vec r,t)] \vec I -\frac{\hbar^2}{4m}n(\vec r,t)\nabla\otimes\nabla \ln n(\vec r,t),
\end{align}
linearization of these equations and Fourier transformation yields the plasmon dispersion for the case of $D=1, 2, 3$ dimensions \cite{zhandos_cpp17_1d}, given by Eq.~(\ref{eq:qhd-dispersion}). 
 \begin{align}
\omega^2_{\rm MQHD}(q) &= \omega^2_{pl}+ \frac{3 v_F^2 q^2}{D+2} + (1-\delta_{2,D})\frac{\hbar^2}{4m^2}q^4,
\label{eq:mqhd-dispersion}\\
\omega^2_{\rm QHD}(q) &= \omega^2_{pl} + \frac{v_F^2 q^2}{D} + \frac{\hbar^2}{4m^2}q^4,
\label{eq:qhd-dispersion}
\end{align}
where we introduced the Fermi velocity via $E_F=mv_F^2/2$. Note that the plasma frequency depends on the dimensionality of the system.

Let us discuss this result. First, we observe that the MQHD-dispersion (\ref{eq:mqhd-dispersion}) exactly coincides with the zero temperature limit of the RPA result presented in Fig.~\ref{fig:DSF_PRL}. In particular, the small-q (large-q) limit given by the first (third) term in Eq.~(\ref{eq:mqhd-dispersion}) are correct. However, comparing with the QMC results (red curves in Fig.~\ref{fig:DSF_PRL}), the MQHD dispersion at intermediate wave numbers, $1 \lesssim q/q_F \lesssim 3$, strongly overestimates the oscillation frequency. This is due to the neglect of  correlation effects in the present calculation. These effects can be restored by including a proper expression for the exchange-correlation potential in the MQHD equations.

Due to the agreement of the MQHD dispersion with the RPA, we can use it to test the accuracy of the QHD model. Note that the QHD model is -- by construction -- less accurate than MQHD because it involves an orbital averaging and thus, a loss of resolution of small length and energy scales. First we observe that the small-q limit is correct. The large-q limit, on the other hand, is correct for 1D and 3D systems, but it is incorrect for 2D quantum plasmas. Third, the behavior at intermediate wave numbers (second term proportional to $q^2$) is correct in 1D. In 2D the QHD yields a coefficient $1/2$, whereas the correct one is $3/4$. Similarly, in 3D the QHD result ($1/3$) deviates from the MQHD result ($3/5$). Therefore, high-frequency electronic plasma oscillations ($\omega \ge \omega_{pl}$) are correctly reproduced only in one-dimensional quantum plasmas. 

Going back to the QHD equations it is easy to verify that the origin of this incorrect coefficient, as compared to the RPA (MQHD) result, is the Fermi pressure term. To recover the correct coefficient of the $q^2$ term in the dispersion (\ref{eq:qhd-dispersion}) the pressure has to be multiplied by a factor $\bar \alpha$ that was reported in Ref.~\cite{zhandos_pop18}: for example for a 3D plasma, the Fermi pressure that appears in Eq.~(\ref{eq:approxpress}) has to be multiplied by a factor $\bar \alpha=9/5$. Note that this value applies only to small wave numbers whereas for large wavenumbers, $q\gg 2k_F$,  it will approach the value $3/5$ \cite{zhandos_pop18}.

\subsection{Screened ion potential from MQHD and QHD. Influence of electronic correlations}\label{ss:qhd-discussion}
\begin{figure}
    \centering
\includegraphics[width=.48\textwidth]{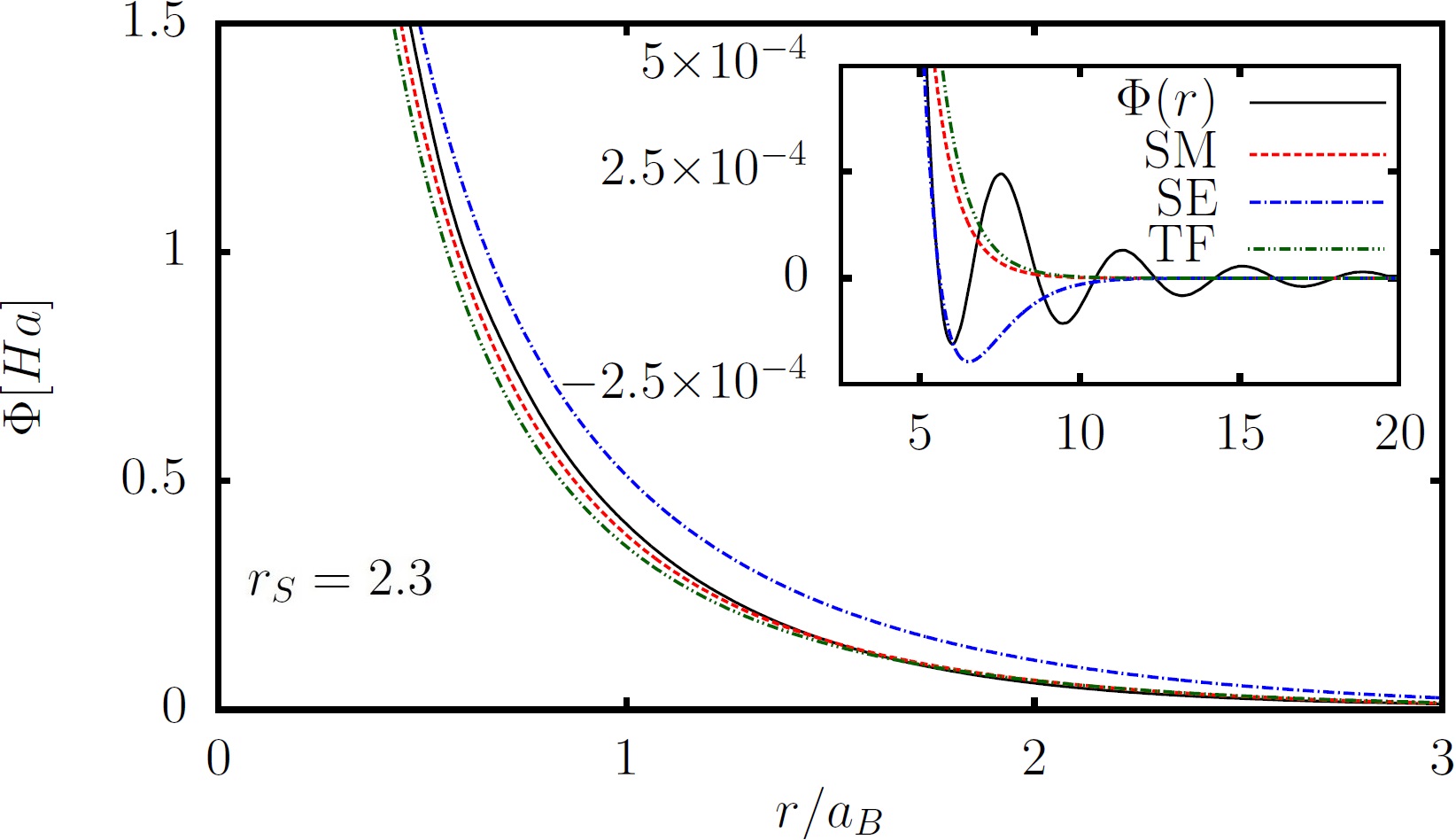}
\vspace{-.2cm}
\caption{Screened potential of a proton in a quantum plasma, for $r_s=2.3$ and $T=0$. Full black line: RPA result showing Friedel oscillations. Red and green lines: potential of Ref.~\cite{stanton_pre_15} and Ref.~\cite{michta_cpp15}, respectively. Blue line (SE): attractive potential of Ref.~\cite{shukla_prl_12} where the Bohm term is used with a prefactor $\gamma=1$ that is in conflict with the MQHD (RPA) result, cf. Fig.~\ref{fig:gamma_theta}. For the effect of electronic correlations, see Fig.~\ref{fig:screened-pot_qmc}. Figure from Ref.~\cite{moldabekov_pop15}, published with the permission of the authors.}
  \label{fig:screened-pot}
\end{figure}

\begin{figure}
    \centering
\includegraphics[width=.35\textwidth]{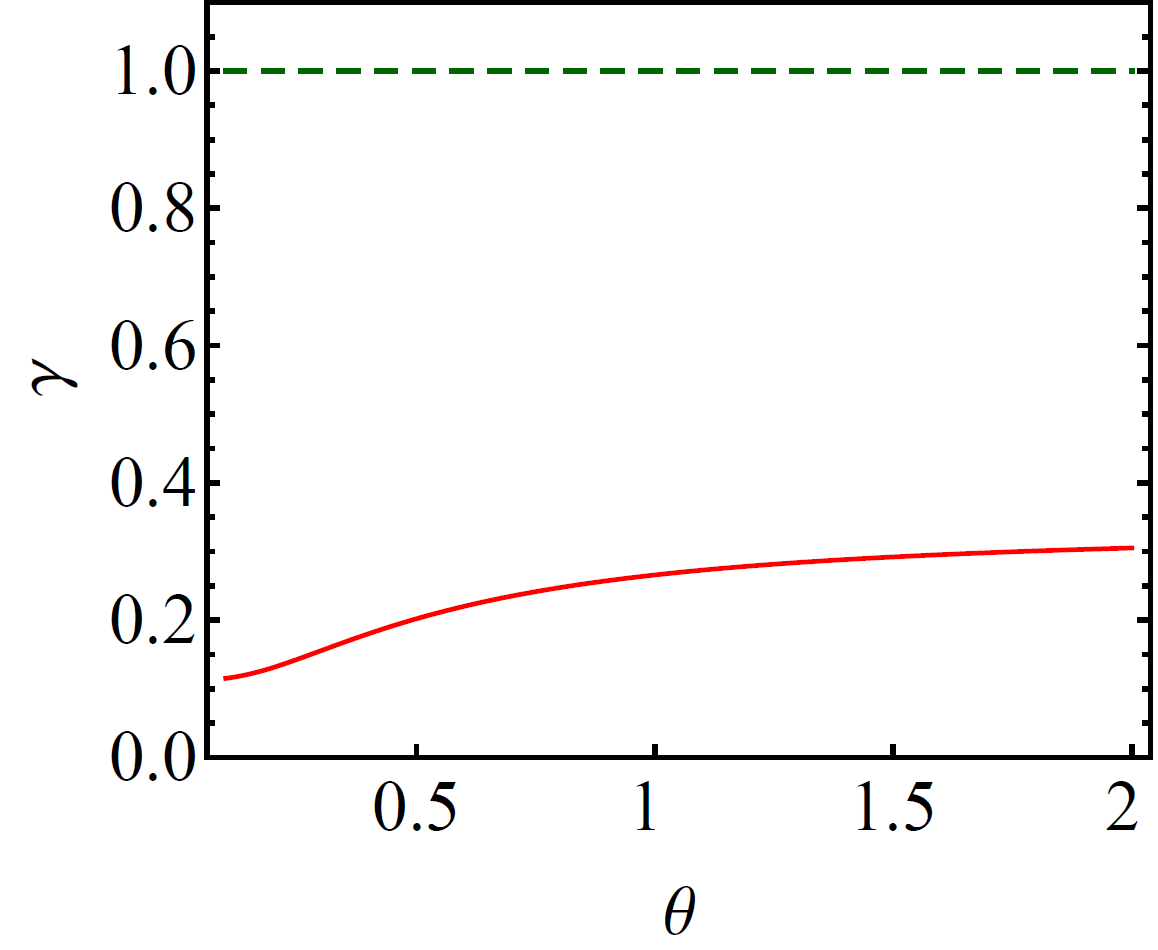}
\vspace{-.2cm}
\caption{Prefactor of the Bohm potential for low-frequency long-wavelength excitations, as a function of the degeneracy parameter. $\gamma$ changes from $1/9$, at $T=0$, to $1/3$, at high temperature. Dashed line: high-frequency limit of $\gamma$. Figure from Ref.~\cite{moldabekov_pop15}, published with the permission of the authors.}
  \label{fig:gamma_theta}
\end{figure}

After having discussed the high-frequency plasma oscillations we now turn to the frequency range $\omega \ll \omega_{pl}$. This is of prime importance, e.g. for ion acoustic oscillations and for the screened potential in a quantum plasma. In fact, screening effects replace the Coulomb potential $Q/r$ by the potential 
\begin{equation} \label{POT_stat}
\Phi(\vec r)   = \int\!\frac{\mathrm{d}^3q}{2 \pi^2 } \frac{Q}{q^2 \epsilon(\vec q, \omega=0)} e^{i \vec q \cdot \vec r} \quad,
\end{equation}
where the dielectric function contains the quantum plasma properties and is taken in the static limit.
The screened potential for quantum plasmas that improves the conventional Yukawa (or Thomas-Fermi) model has been actively studied in recent years, e.g.~\cite{shukla_prl_12,bonitz_pre_13,stanton_pre_15,moldabekov_pop15,akbari_pop_15}. For example, in Ref.~\cite{shukla_prl_12} the authors predicted, using a QHD model that, in a quantum degenerate plasma in thermodynamic equilibrium the electrostatic potential of an ion would be attractive. Their result is shown in Fig.~\ref{fig:screened-pot} by the blue line and exhibits a shallow minimum (depth approximately $5$meV) at about 6 Bohr radii (about 2.5 interparticle distances). Tests with DFT simulations that can be regarded as benchmarks  \cite{bonitz_pre_13,bonitz_pre_13_rep,bonitz_psc_13} revealed that no such minimum exists. The reason for the unphysical predictions of the QHD model was clarified in Ref.~\cite{moldabekov_pop15}: the coefficient of the Bohm term in the QHD equations (\ref{eq:density-balance-fermions_average}) and (\ref{eq:momentum-balance-fermions_average}) turns out to be incorrect for applications to low-frequency excitations. Compared to its value at high frequencies, $\omega\ge \omega_{pl}$ (which we use as a reference, $\gamma=1$), it has to be reduced by almost an order of magnitude ($\gamma \to 1/9$) to reproduce the correct MQHD (RPA) result.
Two screened potentials that contain the correct factor $\gamma=1/9$ are included in the plot as well and do not exhibit the minimum and show good agreement with the full (nonlocal) RPA screened potential \cite{moldabekov_pop15}. The only difference is that these potentials cannot resolve the Friedel oscillations. Note that the correction factor $\gamma$ depends on temperature which is shown in Fig.~\ref{fig:gamma_theta}.

\begin{figure}
    \centering
\includegraphics[width=.49\textwidth]{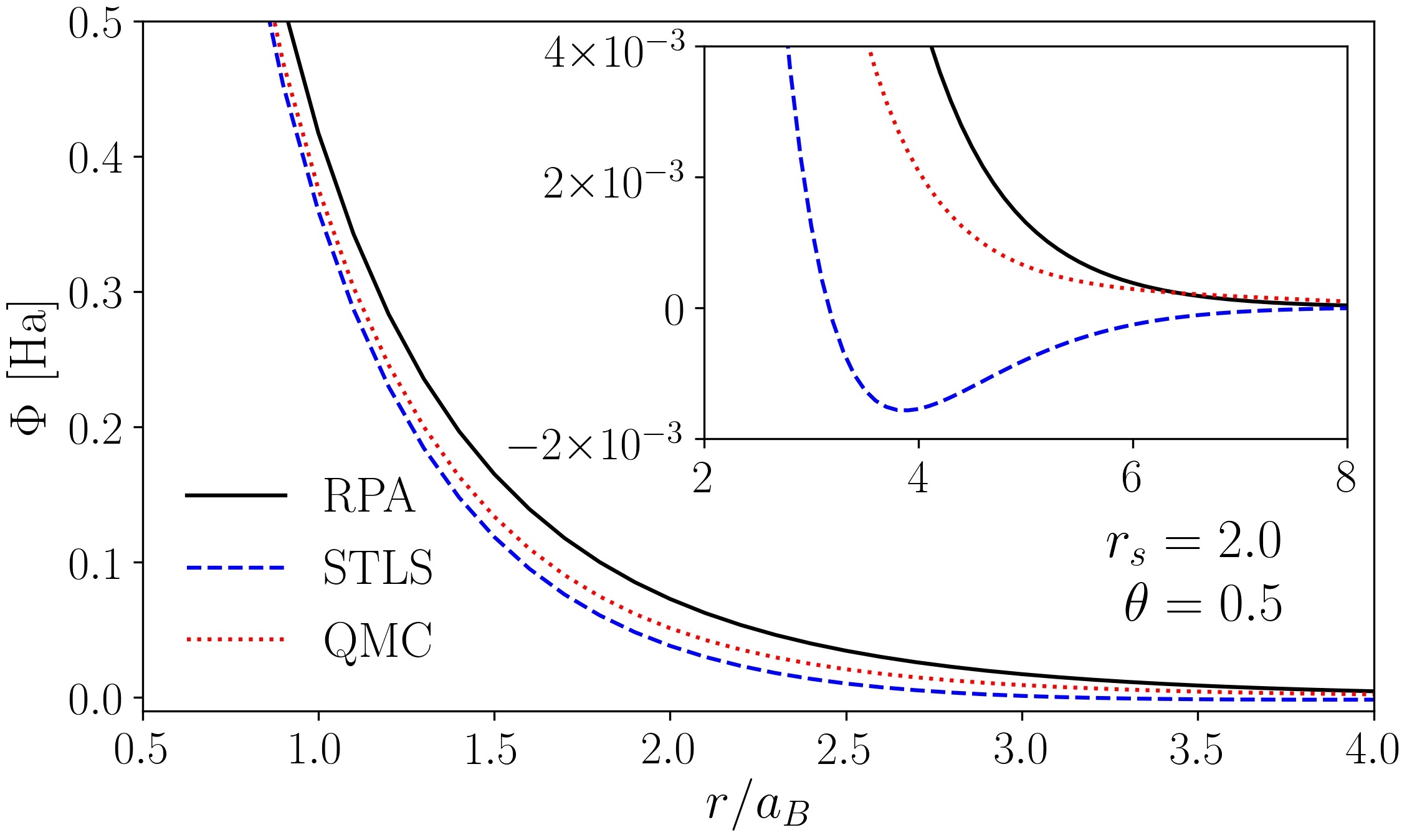}
\vspace{-.2cm}
\caption{Screened potential of a proton in a quantum plasma, for $r_s=2.0$ and $\theta=0.5$, similar to Fig.~\ref{fig:screened-pot}, but here with different electronic local field corrections $G(q)$. Full black line: RPA result ($G(q)\equiv 0$). Blue dashed line: potential computed using $G(q)$ in STLS approximation. Red short-dashed line: $G(q)$ from QMC, using the machine-learning representation [cf. Sec.~\ref{sec:LFC}], presented in Ref.~\cite{dornheim_jcp_19-nn}.}
  \label{fig:screened-pot_qmc}
\end{figure}

Let us now investigate the effect of electronic correlations (effects beyond RPA) on the screened potential.
Here we consider two approximations for the static local field correction $G(q)$ that lead to a correlated dielectric function appearing in formula (\ref{POT_stat}). The first is the standard STLS (Singwi-Tosi-Land-Sj\"olander) approximation and the second, the exact results for $G(q)$ obtained from our QMC simulations, cf. Sec.~\ref{sec:LFC}, where we used the machine learning representation of Ref.~\cite{dornheim_jcp_19-nn}. The two results are shown, together with the uncorrelated RPA result, in Fig.~\ref{fig:screened-pot_qmc}. There we present the screened potential of a proton in a dense quantum plasma in the WDM regime, at $r_s=2$ and $\theta=0.5$ which is close to the parameters of Fig.~\ref{fig:screened-pot}, the main difference being the finite temperature. The first effect of correlations is a significantly stronger screening (more rapid decay of the potential), compared to the RPA case \cite{zhandos_cpp17}. In addition, we observe that the STLS potential strongly deviates from the QMC result not only quantitatively but even qualitatively: it overestimates screening and develops an unphysical attractive part (negative minimum) at intermediate distances (see subplot in Fig.~\ref{fig:screened-pot_qmc})  ~\cite{zhandos_cpp17}. 

As was explored in Ref.~\cite{moldabekov_pre_18}, this unphysical behavior of the STLS approximation leads to additional restrictions on its applicability for two-component plasmas with non-ideal ions. 
Note that, in general, the screening is not exponential~\cite{moldabekov_pre_18} and, in the case of strongly coupled ions, deviations from the RPA screened potential may be quite large and have significant impact on the structural and dynamical properties of the ion component in a dense plasma \cite{zhandos_pre_19, moldabekov_pre_18}. 
Given the overall good accuracy of STLS for thermodynamic quantities, cf. Sec.~\ref{sec:static}, these problems for the screened potential are an unexpected result. This is a similar artefact, as was observed for the QHD screened potential with the wrong Bohm term in Fig.~\ref{fig:screened-pot}. This observation underlines the high importance of \textit{ab initio} QMC results, in particular for the local field correction. The QMC potential presented in Fig.~\ref{fig:screened-pot_qmc} is a preliminary result, and a systematic analysis for a broad range of parameters in the WDM range is an important task of future work.

\subsection{Ion-acoustic modes in a quantum plasma}\label{ss:ion-acoustic}
\begin{figure}
    \centering
\includegraphics[width=.4\textwidth]{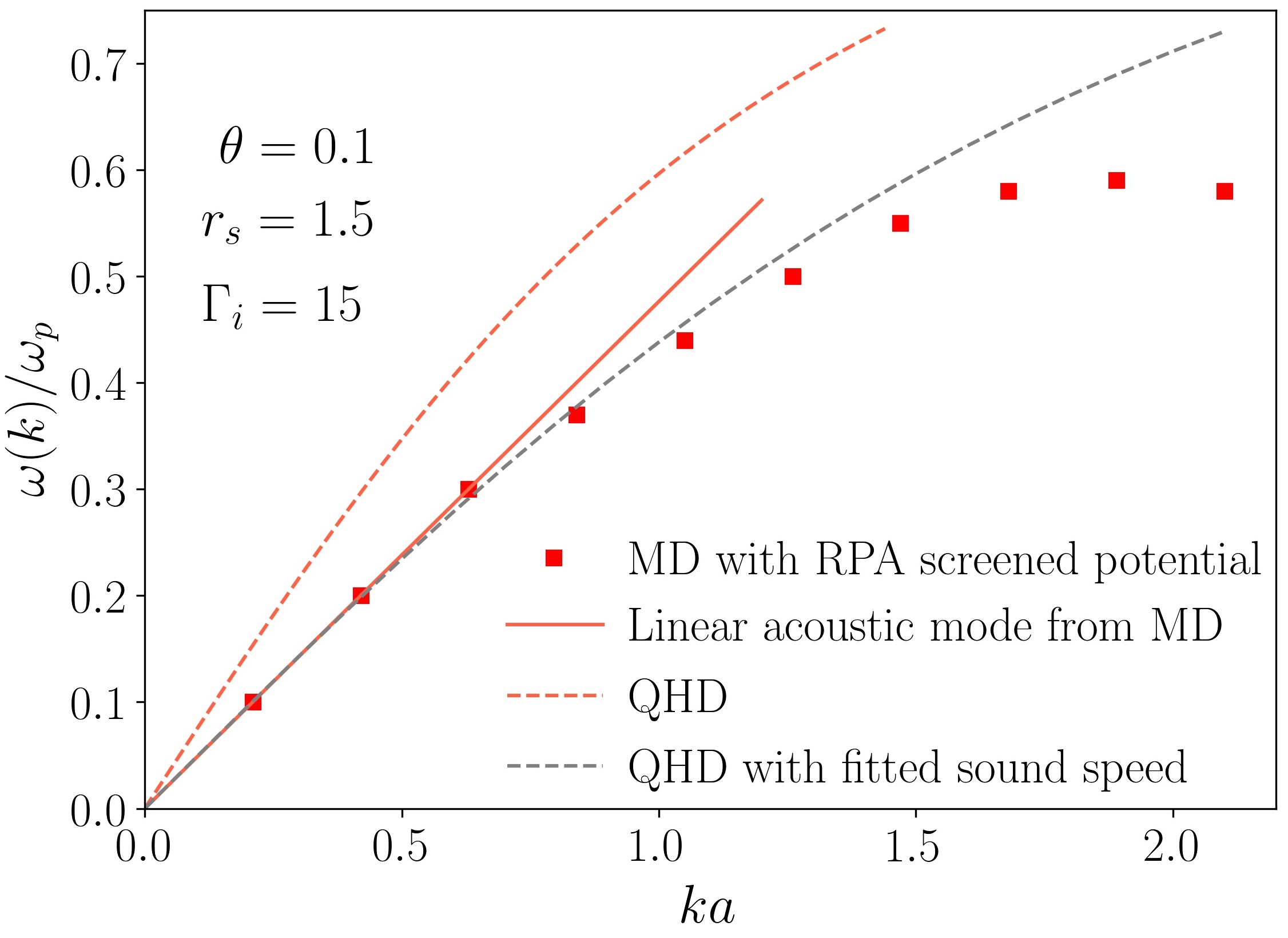}
\vspace{-.2cm}
\caption{Dispersion of ion-acoustic oscillations of nonideal ions (coupling parameter $\Gamma_i=15$) in a weakly correlated quantum plasma ($r_s=1.5, \Theta=0.1$).
Symbols: MD simulation of ions interacting via a screened potential (\ref{POT_stat}) using the RPA dielectric function \cite{zhandos_pre_19}.
Solid straight line: acoustic mode from the MD data; red dashed line: QHD result (\ref{eq:acoustic}) of Ref. \cite{PhysRevE.92.053112} using the sound speed of an ideal quantum plasma; black dashed line: dispersion (\ref{eq:acoustic}) where the sound speed is fitted to the long-wavelength limit of the MD simulations.}
  \label{fig:acoustic_mode}
\end{figure}

Let us now turn to ion-acoustic oscillations in a quantum plasma that have been studied by many authors. 
For example, Schmidt \textit{et al.} gave a hydrodynamic derivation of the dispersion relation and the dynamic structure factor \cite{gregori_pre12}.
Haas and Mahmood \cite{PhysRevE.92.053112} used Euler fluid equations for the ions, neglecting the ionic pressure term, and the QHD equations for the electrons with the Bohm potential with the correct factor $\gamma$,  in the low frequency long-wavelength limit (see Fig.~\ref{fig:gamma_theta}), to study low-frequency waves in a two-component plasma. They derived the following  dispersion for ion-acoustic waves:
\begin{equation}\label{eq:acoustic}
    \omega^2(k)=c_s^2k^2\times\frac{\omega_{pi}^2\left[1+A(\gamma,c_s) k^2\right]}{\omega_{pi}^2+c_s^2k^2\left[1+A(\gamma,c_s) k^2\right]}\,.
\end{equation}
Here $A(\gamma,c_s)=\gamma \hbar^2/(12 m_e m_i c_s^2)$ represents the quantum electron correction due to Bohm term, and $c_s=\left(k_BT/m_i ~Li_{3/2}(z)/Li_{1/2}(z)\right)^{1/2}$ is the ion sound speed of an ideal quantum plasma written in terms of the polylogarithmic function $Li_{\nu}(z)$ of the ideal electron fugacity  $z$. 
In the limit of strong electron degeneracy, the sound speed is given by $c_s=\sqrt{2 E_F/(3 m_i)}$. \\
The dispersion of ion-acoustic wave is sensitive to the properties of the surrounding electrons via screening of the ion potential, Eq.~(\ref{POT_stat}), and to ionic correlations, depending on the coupling parameter $\Gamma_i$. This is illustrated in Fig.~\ref{fig:acoustic_mode} where the QHD result (\ref{eq:acoustic}) is compared to the data from molecular dynamics  (MD) simulation of ions at $\Gamma_i=15$, $r_s=1.5$ and $\theta=0.1$. 
The MD data was obtained using the screened ion potential (\ref{POT_stat}) with $\epsilon(\vec q, \omega=0)$ in RPA \cite{zhandos_pre_19}. As discussed above, in linear response, the RPA (MQHD) description of electrons is more accurate than QHD with the standard Bohm potential and serves as a benchmark. The first observation from Fig.~\ref{fig:acoustic_mode} is that QHD result for the dispersion, Eq.~(\ref{eq:acoustic}), strongly deviates from the MD data for all wave numbers, because the sound speed of an ideal plasma is being used. A much improved behavior is found if the sound speed fitted to the MD data at small wave-numbers is inserted into Eq.~(\ref{eq:acoustic}).  
The resulting dispersion (black dashed curve)
agrees well with the MD data for all wavenumbers,  $ka \lesssim 1.0$. The failure at larger wavenumber is not surprising due to the standard limitation of fluid approaches that we discuss in Sec.~\ref{ss:improved-qhd}.
To summarize, our analysis reveals that the functional form (\ref{eq:acoustic}) of the dispersion is reasonable which is
 promising for further applications of the QHD approach to two-component plasmas, taking into account electronic and ionic non-ideality effects.

\subsection{Limitations and further improvement of QHD: nonlocal and exchange-correlation effects }\label{ss:improved-qhd}
The general validity limits of quantum hydrodynamic models have been discussed in a variety of papers, e.g. Refs.~\cite{manfredi_fields_05,vladimirov_ufn_2011,bonitz_pre_13}. Most importantly, 
similar as in a classical plasma, a sufficient condition for the applicability of a hydrodynamic description is that the considered length scales are larger than the screening length, e.g. \cite{gregori_pre11}. In a quantum plasma this is the Thomas-Fermi length, $\lambda_{TF}$, which leads to the criterion
\begin{align}
    k\lambda_{TF} = \frac{2}{(3\Gamma_q)^{1/2}}\frac{k}{k_F} \ll 1\,.
\end{align}
In the weak coupling limit, $\Gamma_q \le 0.1$, this leads to the restriction $k\lesssim 0.03 k_F$. Previously it was found that the above condition is not always necessary, i.e.  hydrodynamics in many cases applies also to length scales smaller than the screening length, for classical plasmas, see e.g. Ref.~\cite{gregori_pre11}. For correlated quantum plasmas it was found that correlation effects reduce the screening length \cite{moldabekov_pre_18, zhandos_pre_19,zhandos_cpp17}, cf. also Fig.~\ref{fig:screened-pot_qmc}, which supports the same conclusion that QHD should be valid on scales below $\lambda_{TF}$.

Of course, a further extension of the validity range can be achieved if the coefficients in the QHD equations are adjusted using information from the MQHD equations (kinetic theory). For example, it was shown in Ref.~\cite{zhandos_pop18} how to properly choose the QHD parameters for large wavenumber, $k\gtrsim 2k_F$. Moreover, tests against MQHD (RPA) results, discussed in the preceding sections, allow us to correct the prefactors in the QHD equations  
for the high-frequency and low-frequency limits.

Thus, even if correlation effects are neglected ($V^{\rm xc} \to 0,\; J_{np}\to 0,\; Q^\Delta \to 0$), it is clear that there exists no universal result for the parameters in the QHD equations that would apply to arbitrary situations. Instead, depending on the frequency and wave number of the excitation the coefficients in front of the Fermi pressure and of the Bohm term vary, and the results are also dependent on temperature and the system dimensionality:
\begin{eqnarray}
\overline{P}_F &\to& \overline{\alpha}(\omega,q;\Theta,D)\,\overline{P}_F\,,
\label{eq:pf-corrected}
\\
Q_1 &\to& \gamma(\omega,q;\Theta,D)\,Q_1\,.
\label{eq:Q-corrected}
\end{eqnarray}
The values of $\overline{\alpha}$ and $\gamma$ for the important limiting cases of high and low frequency, as well as high and low wave number are known analytically, even at finite temperature \cite{zhandos_pop18}. Thus for these situations reliable simulations are possible. 
The reason for this frequency and wave number dependence of the coefficients is the fact that the kinetic theory (RPA) polarization (density response) function $\Pi^{\rm RPA}$ has different long wavelength limits for different frequencies, $\lim_{q\to 0}\Pi^{\rm RPA}(q,\omega)$. When this is converted into local hydrodynamic equations via orbital averaging, the result is different for high and low frequency, respectively. 

It is possible to avoid this problem by introducing a more general nonlocal expression for the QHD potential that was derived in Ref.~\cite{zhandos_pop18}. Here we summarize these results starting from the RPA and, in addition, including exchange-correlation effects that we link to the dynamic local field correction for which we have obtained \textit{ab initio} results via QMC simulations, cf. Sec.~\ref{sec:LFC}.
The main idea behind non-local quantum hydrodynamics is to    require \cite{zhandos_pop18} that the QHD polarisation function equals to the polarisation function that follows from kinetic theory or TD-DFT (MQHD), i.e  $\Pi_{\rm QHD}\equiv \Pi_{\rm LR}$. The derivation of the QHD equations presented in Sec.~\ref{ss:averaging} gives a strict justification for this requirement.

Let us return to the QHD momentum equation (\ref{eq:momentum-balance-fermions_average}), considering zero vorticity, and introduce the total potential $\mu[n(\vec r ,t)]$ 
that contains ideal and exchange-correlation contributions (first and second terms),
\begin{eqnarray}\label{eq:QHD_momentum_POP18}
&&\frac{\partial \overline{\textbf{p}}}{\partial t} + \frac{1}{m}(\overline{\textbf{p}}\cdot
\nabla)
\overline{\textbf p} = -\nabla \mu[n(\vec r ,t)], \quad \\
&& \mu[n(\vec r ,t)]=\mu_{\rm id}[n(\vec r ,t)]+\mu_{\rm xc}[n(\vec r ,t)] +e\phi(\vec r,t),
\label{eq:QHD_mu_POP18}
\end{eqnarray}
whereas the last term is due to the total field -- the sum of external field as well as mean field (Hartree) contributions which is the solution of Poisson's equation, i.e. $e\phi = U_{\rm tot}$. %
The force field $\mu$ is defined by the functional derivative of the grand potential  \cite{zhandos_pop18}:
\begin{align}
    \mu[n(\vec r ,t)]-e\phi(\vec r,t) &=\frac{\delta \Omega[n(\vec r,t)]}{\delta n(\vec r,t)} \,,
    \label{eq:QHD_potential_POP18}
\end{align}
where $\Omega= \Omega_{\rm id}+\Omega_{\rm xc}$, is the sum of an ideal and exchange-correlation part that will be specified below. 

In equilibrium (current-free case), Eqs.~(\ref{eq:QHD_momentum_POP18}) and (\ref{eq:QHD_potential_POP18}) reduce to the Euler-Lagrange
equation 
\begin{equation}
 \frac{\delta \Omega[n_0(\vec r,t)]}{\delta n_0(\vec r,t)} +e\phi_0(\vec r,t)=0, 
\end{equation}
where the subscript ``$0$'' indicates the equilibrium case.
Assuming a weak perturbation \cite{1974_Ying, zhandos_pop18}, the force becomes
 \begin{multline} \label{eq:d2T_new}
\nabla \mu[n(\vec r ,t)]=\nabla \int \mathrm{d}\vec{r}^{\prime}\, \left.\frac{\delta^2 \Omega[n]}{\delta n(\vec{r},t)\delta n(\vec{r^{\prime}}, t)}\right|_{\rm n=n_0} n_1(\vec{r}^{\prime},t),
  \end{multline} 
  where $n_1(\vec r, t)=n(\vec r, t)-n_0(\vec r)$, with $|n_1/n_0|\ll1$.

Equations~(\ref{eq:QHD_momentum_POP18})-(\ref{eq:d2T_new}) and the requirement that the QHD density response agrees with the kinetic theory results in linear response,  $\Pi_{\rm QHD}\equiv \Pi_{\rm LR}$, give \cite{zhandos_pop18}:
 \begin{align}\label{eq:grand_potential}
\mathfrak{F}\left[\left.\frac{\delta^2 \Omega}{\delta n(\vec{r}, t)\delta n(\vec{r^{\prime}}, t)}\right|_{\rm n=n_0}\right]
&= -\frac{1}{\Pi'_{\rm LR} (k, \omega)},
\\
\frac{1}{\Pi'_{\rm LR} (k, \omega)} &= 
\frac{1}{\Pi_{\rm LR} (k, \omega)} - \frac{1}{\Pi^0(\omega)}\,,
\end{align} 
where we introduced the modified linear response polarization, $\Pi'_{\rm LR}$ from which the long-wavelength limit at finite $\omega$, $\Pi^0(\omega)= m k^2/\omega^2 n_0$, is being subtracted, 
 and $\mathfrak{F}$ denotes the Fourier transform to frequency and wavenumber $(k, \omega)$ space. 
Considering first the case of ideal quantum electrons, where $\Pi_{\rm LR}=\Pi_{\rm RPA}$,  we have:
 \begin{multline}\label{eq:ideal_grand_potential}
\mathfrak{F}\left[\left.\frac{\delta^2 \Omega_{\rm id}}{\delta n(\vec{r}, t)\delta n(\vec{r^{\prime}}, t)}\right|_{\rm n=n_0}\right]
= -\frac{1}{\Pi'_{\rm RPA} (k, \omega)}\,.
\end{multline} 
Equation~(\ref{eq:ideal_grand_potential}) yields a non-local potential, Eq.~(\ref{eq:d2T_new}), and allows, among others, to find the correction factors $\overline{\alpha}(\omega,q;\Theta,D)$ and $\gamma(\omega,q;\Theta,D)$, Eqs.~(\ref{eq:Q-corrected}) and (\ref{eq:pf-corrected}), in various limiting cases \cite{zhandos_pop18}.

Consider now the general case of nonideal electrons, $r_s\gtrsim 1, \Gamma_i\gtrsim 1$, cf. Sec.~\ref{s:qp_parameters}. This requires to  include the exchange-correlation contribution to the grand potential which can be done by using approximations for 
the exchange-correlation potential, $V_{\rm xc}$, from DFT. A simple ground state expression in the local density approximation was introduced to QHD in Ref. \cite{manfredi_prb_08}) and has been used in a number of subsequent papers, e.g. in Ref.~\cite{El-Labany2010}. 
A second strategy that is closer to the topics discussed in this paper is to directly use \textit{ab initio} input from QMC. This is indeed possible via the dynamic local field correction $G(q,\omega)$, that was computed in Sec.~\ref{ss:dsf}, \cite{zhandos_pop18}:
  \begin{equation}\label{eq:Omega_xc}
\mathfrak{F}\left[\left.\frac{\delta^2 \Omega_{\rm xc}}{\delta n(\vec{r})\delta n(\vec{r^{\prime}})}\right|_{\rm n=n_0}\right]=-\tilde v_k
G(k,\omega).
  \end{equation} 
  Equations (\ref{eq:QHD_momentum_POP18})-(\ref{eq:QHD_potential_POP18}), (\ref{eq:ideal_grand_potential}) and (\ref{eq:Omega_xc}) represent a closed set of equations which is exact in the weak perturbation case, i.e. $\left|n_1/n_0\right|\ll1$, and can be summarized in one generalized nonlocal momentum balance equation \cite{zhandos_pop18}:
  \begin{widetext}
\begin{multline}
 \frac{\partial  \overline{\textbf{p}}(\vec{r},t)}{\partial t}=-\nabla e\varphi_1(\vec{r}, t)
-\nabla \int \mathrm{d}\vec{r}^{\prime}\,n_1(\vec{r}^{\prime},t)
\pmb{\left[\vphantom{\frac{1}{2}}\right.}
\int   \frac{\mathrm{d}\vec{k}}{(2\pi)^3}\, \mathrm{d} \omega\, e^{i[\vec k\cdot(\vec r-\vec r^{\prime})-\omega t]}\left(-\frac{1}{\Pi'_{\rm RPA} (k, \omega)}
- \tilde v_k
G(k, \omega)\right)
\pmb{\left.\vphantom{\frac{1}{2}}\right]}
\bigg|_{\rm n_0},
\label{eq:generalized_qhd}
\end{multline}
where $\varphi_1(\vec r, t)=\varphi (\vec r, t)-\varphi_0(\vec r)$\,.
\end{widetext}
This is a remarkable result that contains all relevant limiting cases and assures the highest accuracy possible via a link to quantum kinetic theory and \textit{ab initio} input from QMC.
  
Let us discuss the limitations of the result (\ref{eq:generalized_qhd}). The main assumption in the derivations above is the validity of linear response, i.e.  $\left|n_1/n_0\right|\ll1$. If this is not satisfied, equations
(\ref{eq:QHD_momentum_POP18})-(\ref{eq:QHD_potential_POP18}), (\ref{eq:ideal_grand_potential}) and (\ref{eq:Omega_xc}) can still be used, but the accuracy will be largely defined by the form of the functional $\Omega [n]$, e.g. see \cite{witt_del, Sjostrom_PhysRevB.88},  and the results will be not reliable. This concerns, in particular, applications to nonlinear oscillations and waves in quantum plasmas. In this case,
the linear response result, $\Pi_{\rm LR}$, has, in principle, to be replaced by solutions of a nonlinear kinetic equation.   
Another case that is beyond the scope of Eq.~(\ref{eq:generalized_qhd}) is very rapid external excitation. In this case the distribution function may be far from a Fermi function $f^{\rm eq}$, giving rise to a strongly modified plasmon spectrum in quantum plasmas, e.g. \cite{bonitz-etal.93prl,bonitz_prb_0}, similar to the case of classical plasmas. The relevance of nonequilibrium plasmas under warm dense matter conditions was studied, e.g. by Gericke \textit{et al.} \cite{gericke_prl11}. For these situation, a nonequilbrium quantum kinetic theory is required that yields the time-dependent density response, $\epsilon^{\rm eq}(q,\omega;t)$ \cite{vorberger_pre18} and local field corrections, $G^{\rm eq}(q,\omega;t)$ which assumes an equilibrium form in which the distribution function is replaced by a nonequilibrium function, $f^{\rm eq} \to f(t)$. However, even this approach maybe inappropriate if the excitation is on the scale of the plasma period or faster, $t\lesssim 2\pi/\omega_{pl}$. In that case, the formation of the screening cloud and of the plasmon spectrum proceeds on the time scale of the excitation, and  a true nonequilibrium theory for the density response functions is required, cf. Ref.~\cite{bonitz_qkt} and references therein.

\section{Conclusions and outlook} \label{s:conclusion}
In this article we have presented an overview on recent simulation results for warm dense matter. First, we have presented thermodynamic results for the degenerate electron component, considering the warm dense uniform electron gas, that are based on \textit{ab initio} quantum Monte Carlo simulations. The results include highly accurate parametrizations of the exchange-correlation free energy and results for the static local field correction $G(q)$. Furthermore, we presented \textit{ab initio} results for dynamic quantities, including the dynamic local field correction $G(q,\omega)$, the dynamic structure factor $S(q,\omega)$ and the dynamic density response $\chi(q,\omega)$. These results can be further extended to other dynamic quantities that are of high relevance for current warm dense matter experiments. Moreover, the static and dynamic local field correction are of key importance as input for other models and  simulations methods, including quantum hydrodynamics and density functional theory. 
\begin{figure}
    \centering
\includegraphics[width=.48\textwidth]{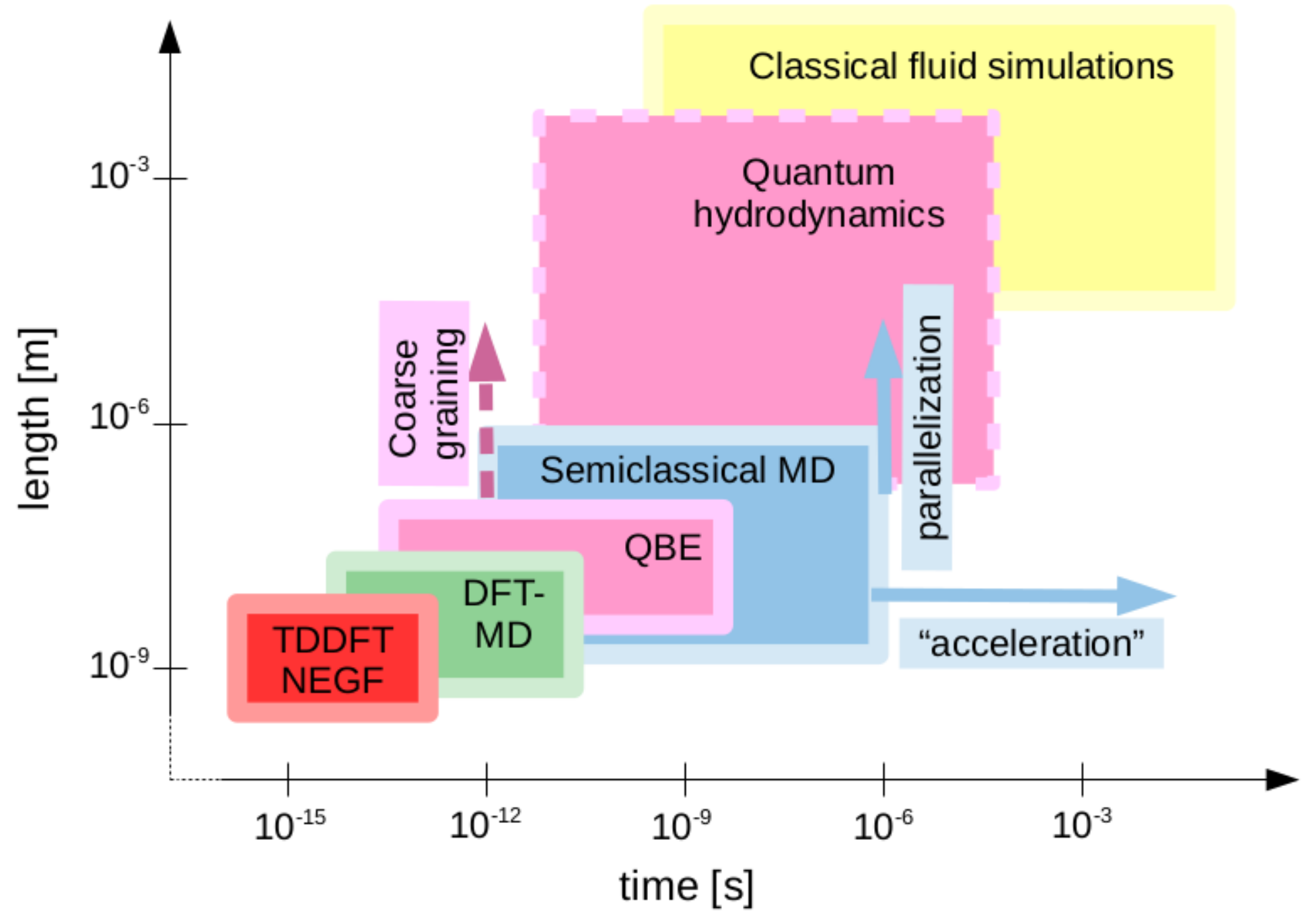}
\vspace{-.2cm}
\caption{Overview on simulation methods of relevance for warm dense matter and their approximate length and time scales. DFT-MD: Born-Oppenheimer density functional theory; TDDFT: time-dependent DFT;
QBE: quantum Boltzmann equation, NEGF: Nonequilibrium Green functions. Figure modified from Ref.~\cite{Bonitz_fcse_19}.}
  \label{fig:methods_space-time}
\end{figure}
DFT today is the only approach that is, eventually, capable to cover the entire WDM range including electron-ion plasmas as well as the condensed matter phase. These simulations give access to time-dependent properties: in Born-Oppenheimer MD the electrons adiabatically following the ions.
In contrast, in time-dependent DFT a real (non-adiabatic) time-dependence of electrons and ions is described.

However, DFT is notoriously inaccurate in treating ionization energies, band gaps and electronic correlation effects, in particular under WDM conditions. Here, \textit{ab initio} input from QMC can substantially improve the simulations. One such input is the exchange-correlation free energy that has been used to improve the local density approximation, cf. Sec.~\ref{sec:static}. While the finite-temperature effect on the LDA equation of state turned out to be just on the order of a few percent, this seems to bring the results significantly closer to QMC simulations of dense plasmas, cf. Sec.~\ref{s:dft}. Even more promising are the QMC data for the static and dynamic local field correction that provide highly valuable input for improved exchange-correlation kernels of TD-DFT and QHD, cf. Secs.~\ref{ss:dsf} and \ref{ss:improved-qhd}.

At the same time, the present versions of DFT and TD-DFT are not able to describe thermalization and electronic correlation effects, such as Auger processes and double excitation. These processes require a kinetic description in the frame of quantum kinetic theory or nonequilibrium Green functions (NEGF). As an example we mention the modeling of laser-plasma interaction, including inverse bremsstrahlung heating, harmonics generation, e.g. \cite{kremp_99_pre,haberland_01_pre,bonitz_99_cpp}, and laser absorption during shock compression of matter \cite{zhao_PhysRevLett.111.155003}.  However, both NEGF and TD-DFT are extremely computationally costly and, therefore, are currently limited to short length and time scales. Larger length and time scales are accessible with simpler approaches, such as DFT-MD [cf. Sec.~\ref{s:dft}] or the quantum Boltzmann equation \cite{bonitz_qkt}. A qualitative summary of the length and time scales that can be described by the different methods is given in Fig.~\ref{fig:methods_space-time}. There we also include molecular dynamics with quantum potentials (semiclassical MD) which is applicable when the electronic quantum dynamics are not important (for additional simulation approaches and references see the Introduction). We also indicated the possibility to extend these simulations to larger length scales by means of parallelization, whereas longer time scales can only be achieved, in some cases, by acceleration concepts, e.g. by combination with analytical models, for a discussion see Refs.~\cite{Bonitz_fcse_19,bonitz_psst18,filinov_psst18_1}.

Thus, there is still a big gap between the length and time scales that are accessible by microscopic simulations and the scales that are relevant in experiments. Here, 
a reliable fluid theory for fermions that is similarly advanced and successful as in classical plasmas such as the QHD studied above, could serve as the missing link. As we have pointed out, the orbital averaging involved in deriving the QHD equations restricts this model to finite length and time scales. This means, fast processes, in particular related to the thermalization of the electron distribution, as well as small length scales on the order of the Bohr radius or the Thomas-Fermi screening length, cannot be resolved. For these effects more advanced methods, such as DFT and quantum kinetic theory have to be used (see below). At the same time, the relative simplicity of the QHD equations allows one, to extend them to large length scales and propagate them to long times where the aforementioned approaches cannot be appliied. Thus, QHD could be a very useful tool that is complementary to other methods. This situation is indicated qualitatively  in Fig.~\ref{fig:methods_space-time}.

  \section*{Acknowledgments}
    We thank F. Graziani for useful discussions.
SZ acknowledges financial support from the German Academic Exchange Service (DAAD) together with China Scholarship Council (CSC) and funding from National Natural Science Foundation of China (NSFC Grant No. 11904401).
   This work has been supported by the Deutsche Forschungsgemeinschaft via grants BO1366/13 and BO1366/15 and by the HLRN via grants shp00023 and shp00015 for computing time. Some computations were performed on a Bull Cluster at the Center for Information Services and High Performance Computing (ZIH) at TU Dresden. We would like
to thank the ZIH for its support and generous allocations of computer time.
   TD acknowledges support by the Center for Advanced Systems Understanding (CASUS) which is financed by Germany's Federal Ministry of Education and Research (BMBF) and by the Saxon Ministry for Science and Art (SMWK) with tax funds on the basis of the budget approved by the Saxon State Parliament.
   ZM acknowledges support by    the Ministry of Education and Science of Kazakhstan under Grant No. BR05236730.

%


\end{document}